\documentclass{aa}  
\usepackage{graphicx}
\usepackage{lscape}                 	 
\usepackage{txfonts}
\newcommand{\Teff} {T$_{\rm eff}$}
\newcommand{\grav} {log\,{\em g}}
\newcommand{\vsini} {$v$\,sin\,$i$}
\newcommand{\kms} {km\,s$^{-1}$}
\newcommand{\fwhm} {{\sc fwhm}}
\newcommand{\ft} {{\sc ft}}
\newcommand{\fastwind} {{\sc Fastwind}}
\newcommand{\idl} {{\sc idl}}
\newcommand{\snr} {{\sc SNR}}
\newcommand{\linea}[3]{{\ion{#1}{#2}\,$\lambda$\,#3}}
\newcommand{\solu}[2]{#1\,$\pm$\,#2}
\newcommand{\macro}{$v_{\rm m}$}
\begin{document}
   \title{Fourier method in the determination of rotational velocities in OB stars}


\author{S. Sim\'on-D\'{\i}az\inst{1,2} and A. Herrero\inst{1,3}}

\institute{Instituto de Astrof\'\i sica de Canarias, E-38200 La Laguna, 
           Tenerife, Spain
		   \and
		   LUTh, Observatoire de Meudon, 92195 Meudon Cedex, France
           \and
           Departamento de Astrof\'{\i}sica, Universidad de La Laguna, 
		   E-38071 La Laguna, Spain}

\offprints{sergio.simon-diaz@obspm.fr}
\date{Submitted/Accepted}

\titlerunning{FT method in OB stars}
\authorrunning{Sim\'on-D\'{\i}az \& Herrero}

 
  \abstract
   {}
   {We present a comprehensive study that applies the Fourier transform 
to a sample of O and early B-type stars (either dwarfs, giants, or 
supergiants) to determine their projected rotational velocities, compare 
with previous values obtained with other methods, and seek for evidence 
of extra broadening in the spectral lines}
   {The Fourier technique, extensively used in the study of cooler stars, 
   has remained only marginally applied for the case of early-type stars.
   The comparison of \vsini\ values obtained through the \ft\ and \fwhm\ 
methods shows that the \fwhm\ technique must be used with care in the analysis 
of OB giants and supergiants, and when it is applied to \ion{He}{i} lines. 
Contrarily, the \ft\ method appears to be a powerful tool to derive reliable 
projected rotational velocities, and separate the effect of rotation from other 
broadening mechanisms present in these stars.}
   {The analysis of the sample of OB stars shows that while dwarfs and giants 
display a broad range of projected rotational velocities, from less than 30 up 
to 450 \kms, supergiants have in general values close to or below 100 \kms. 
The analysis has also definitely shown that while the effect of extra broadening 
is negligible in OB dwarfs, it is clearly present in supergiants.
When examining the behavior of the projected rotational velocities with the 
stellar parameters and across the HR diagram, we conclude, in agreement with 
previous researchers, that the rotational velocity should decrease when the 
stars evolve. On the contrary, macroturbulence may be constant, resulting 
therefore in an increasing importance as compared to rotation when the stars 
evolve.}
   {}


   \maketitle

\section{Introduction}
\label{section1}
Together with mass and metallicity, stellar rotation (i.e. stellar angular 
momentum) determines the formation, structure and evolution of a star. Moreover, 
it is a necessary property for a large number of important processes, like the 
solar dynamo or the stellar activity. 
In the realm of massive stars, rotation is becoming increasingly important. It 
is thought to be the reason for internal mixing mechanisms that would bring CNO 
processed material and transport inner angular momentum to the stellar surface 
(Maeder \& Meynet \cite{Mae00}). Coupled with stellar winds, it can strongly 
alter the evolution of the star and the theoretical Eddington limit. 
Equally important, the coupling with stellar winds imposes a dependence of the 
rotational velocity itself with metallicity during the evolution of high mass 
stars (Maeder \& Meynet \cite{Mae01}). Consequently, massive stars in the 
Magellanic Clouds are expected to have larger rotational velocities in average, 
and more CNO contamination in their atmospheres. 

Since present evolutionary stellar models take this rotational velocity into 
account, considering it as an important parameter in the stellar evolution, 
accurate determination of stellar rotational velocities in massive stars is 
required, not only to constrain these stellar evolutionary models, but also to 
understand the physics of massive stars and their dependence with metallicity.\\

Several methods for the determination of the projected rotational velocity
(\vsini) in isolated early-type stars can be found in the literature; the 
most commonly used are:
\begin{enumerate}
\item {The Full-Width-Half-Maximum (\fwhm) of the observed line 
profile is measured and related to the \vsini\ (\fwhm\ method).}
\item {The observed spectrum is cross-correlated against a 
{\em template} spectrum of low rotational velocity. The Gaussian width of the 
cross-correlation function is directly related to the projected rotational 
velocity of the star if rotation is the dominant broadening mechanism of the 
photospheric lines.}
\item {The observed line profiles are compared with synthetic 
ones calculated from model atmospheres that are convolved with the    
corresponding broadening functions. Although this is a very accurate method 
if the broadening functions are well known, too much computational time is 
required. Reducing that time (e.g., using model grids) reduces also 
the accuracy.}
\item {The Fourier transform of the line profile is calculated. 
The position of zeroes present in this Fourier transform depends on the 
\vsini.}
\end{enumerate}
Being very fast and easy to use, the \fwhm\ method has been extensively
used for the measurement of projected rotational velocities in OB stars
(viz. Slettebak  \cite{Sle75}, Abt et al. \cite{Abt02}, Herrero et
al. \cite{Her92}, Strom et al. \cite{Str05}). The cross-correlation method 
was applied by Penny (\cite{Pen96}), Howarth et al. (\cite{How97}) and Penny 
et al. (\cite{Pen04}) to {\sc iue} and {\sc stis} spectra of O-type stars 
and early B-type supergiants.
The studies coming from both methodologies (i.e., \fwhm\ and cross-correlation)
allow us to have \vsini\ measurements of large samples of OB stars with 
different luminosity classes; however, the main limitation of these techniques 
is that they do not allow for the separation of rotational broadening from 
other possible extra broadening mechanisms, usually referred to as 
macroturbulence (although the actual physical mechanism - or mechanisms -
remains unknown). 
In fact, this is a delicate and 
quite important task, since many works find an absence of narrow-lined OB stars. 
This result has been claimed to be an indirect proof of the presence of a 
non-negligible line broadening mechanism in addition to rotation in these 
stellar objects, specially in the case of giants and supergiants (viz. 
Slettebak \cite{Sle56}, Conti \& Ebbets \cite{Con77}, Howarth et al. 
\cite{How97}).

The fitting to broadened synthetic profiles makes it possible to distinguish 
between different broadening mechanisms and also to determine other stellar 
parameters, but as has been commented before, a large effort is required 
to use it accurately.
Furthermore, quite good quality spectra (i.e. high signal-to-noise ratio,
\snr, and spectral resolution) are required to disentangle the various 
broadening mechanisms that may affect the line profiles, and the derived 
values may depend on the considered models and the definition of the 
broadening functions. An early attempt to separate the rotational and 
macroturbulent contributions to the line broadening in O- and early B-type 
stars was performed by Slettebak  (\cite{Sle56}), however he found that 
the quality of his spectra was not good enough to get any conclusive 
solution.

More recently,  Ryans et al. (\cite{Rya02}) analyzed high quality spectra 
of twelve B-type supergiants. They compare observed and synthetically 
broadened lines using a $\chi^2$ technique and find for their survey of 
B-type supergiants that while models dominated by rotation provide 
unsatisfactory fits, those where macroturbulence dominates and rotation 
is negligible (as a broadening agent) were acceptable.

Being also very fast and easy to use, the Fourier method, a technique  
that was pioneered using cool stars, has been only marginally applied in 
the study of OB stars. This technique allows the \vsini\ to be easily 
derived independently of any other broadening mechanism affecting the 
line profile. There are numerous papers illustrating its use in 
the determination of \vsini\ in later type stars (viz. Smith \cite{Smi76a}, 
Gray \cite{Gra80}, Royer et al. \cite{Roy02a}, \cite{Roy02b}, Reiners et 
al. \cite{Rei03}); however, there is no comprehensive study  
applying this technique to early-type stars except that by Ebbets 
(\cite{Ebb79}), who analyzed a sample of 16 objects, but the limited quality
of the spectra allowed him to obtain estimations of the macroturbulence
only for half the objects in the sample. Recently, Sim\'on-D\'{\i}az et 
al. (\cite{Sim06}) have applied the Fourier method to a small sample of 
OB stars in the Orion nebula. The strength of this method was shown in 
the study of the main 
ionizing star of this nebula, an O7\,V star with a well known rotational 
period of $\sim$\,15 days, that is in conflict with the \vsini\ derived 
from the \fwhm\ method, but agrees with the Fourier technique. 

Following the ideas of the present paper and Ryans et al. (\cite{Rya02}), 
Dufton et al. (\cite{Duf06}) have recently analyzed the high-resolution 
spectra for 24 SMC and Galactic B-type supergiants to estimate the 
contributions of both macroturbulence and rotation to the broadening 
of their metal lines. See also the analysis of periodically variable
B supergiants by Lefever et al. (\cite{Lef06}).\\

This paper has three main objectives. First, we extend the work presented in 
Sim\'on-D\'{\i}az et al. (\cite{Sim06}) illustrating the strength of the 
Fourier method in the analysis of early type stars; second, we provide new, 
more accurate values for the rotational velocities of OB stars and 
estimate the role played by other broadening mechanisms; and third, we 
extend the work by Dufton et al. 
(\cite{Duf06}) and Lefever et al. (\cite{Lef06}) towards early spectral 
types. 

We restored from our archive the spectra of several OB stars, obtained in 
different observing campaigns, for a comprehensive determination of the 
rotational velocities through the Fourier analysis method. But firstly, a 
formal study was done to determine the range of applicability of this 
methodology to OB stars.\\ 

The paper is structured in four main parts as follows: a short review 
of the Fourier methodology, as well as some formal tests for the case of 
massive stars are presented in Sect. \ref{section2}. The determination 
of the projected rotational velocities of the sample of OB type stars is 
presented in Sect. \ref{section3}. Finally, the discussion and the main 
conclusions are presented in Sects. \ref{section4} and \ref{section5}.
%
\section{Testing the Fourier method in early-type stars}
\label{section2}
%
The theoretical base of this methodology was well established by Gray 
(\cite{Gra76}; see also the latest edition of the book, published in 
\cite{Gra05}). 
Here, we briefly summarize the main characteristics of the method, needed to 
understand its advantages and limitations, and to evaluate the results 
presented in Sect. \ref{section3}. We also present examples specifically aimed 
at its application to early-type stars. For a more precise description
of the methodology and its applications to late-type stars, we also refer 
to Gray's book (see above for references). Additional details may be found 
in the pioneering articles by Gray (\cite{Gra73}, \cite{Gra75}), Smith 
(\cite{Smi76a}), and Smith \& Gray (\cite{Smi76b}). 
Some objections to its validity in special cases can be found in Mihalas 
(\cite{Mih79}) and Bruning (\cite{Bru84}).\\

In a simple way, an observed stellar line profile may be considered to 
consist of the convolution of an intrinsic profile with the rotational, 
macroturbulent and instrumental profiles. The intrinsic profile includes 
natural and thermal broadenings, as well as other contributions, as those 
due to microturbulence and Stark effect.
The Fourier method (\ft) for the determination of \vsini\ (c.f. Gray 
\cite{Gra76}) is based on the fact that between the previous broadening 
profiles, only the rotational function is expected to have zeroes in its 
Fourier transform (except for large values of microturbulence; see below). 
As firstly described by Carroll (\cite{Car33}), the 
position of these zeroes in frequency space depends on the \vsini\ of the 
star, so that the frequency of the first zero ($\sigma_{1}$) is related 
to the rotational velocity through:
%
\begin{equation}\label{form1}
\frac{\lambda}{c}\,v\,$sin$\,i\, \sigma_{1} = 0.660
\end{equation}
%
Since convolutions transform into products in Fourier space, these zeroes 
will also appear in the total transform function of the line profile. 
Therefore, it is in principle possible to determine, independently 
of any other broadening mechanism, the \vsini\ of a star once the position 
of the first of those zeroes is identified in the Fourier transform of the 
line profile\footnote{This argument is only valid under the hypothesis 
considered by the convolution method, which assumes that the effects of the 
different broadening mechanisms act independently to broaden the emergent 
flux profile (Mihalas \cite{Mih79}).}.

Other mechanisms may contribute with zeroes to the total Fourier 
transform, depending on their profiles in the wavelength domain. Note, for 
example, that microturbulence can also add zeroes, when strong lines are 
considered, due to the boxiness caused by it in saturated lines (Gray 
\cite{Gra76}). However, those zeroes arising from saturation in the Fourier 
space appear at higher frequencies than those for rotation, so there is 
rarely any confusion between them, except for those cases with low \vsini.
%
\subsection{Some formal tests for massive stars}
\label{section21}
%
The properties and advantages of the Fourier method, particularly to 
distinguish among physical processes that produce very similar profiles in 
the wavelength domain, have been already discussed and illustrated in the 
works mentioned above. Here we present a few applications to the particular 
case of early-type stars.

For this formal tests we used \fastwind\ (Santolaya-Rey et al. \cite{San97}, 
Puls et al. \cite{Pul05}) theoretical profiles from a model with 
\Teff\,=\,30000 K, \grav\,=\,4.0 (an early B dwarf) convolved with different 
broadening profiles. Thermal and microturbulence broadenings were included 
in the line calculation in the usual way. 

For the rotational broadenings we considered the usual formulation 
adopted for the rotational broadening (Gray \cite{Gra76}) which assumes a 
rotational profile that is convolved with the flux profile. As shown by
Conti and Ebbets (\cite{Con77}) this approximation may produce a small
overstimation
of the derived rotational velocity of fast rotators. However, as shown by these
authors, the effect is negligible below 200 km/s and remains small above that
value.

The treatment of macroturbulence is a more delicated issue. Several 
authors have pointed out towards indirect evidences of some type of 
macroturbulent broadening acting in the case of OB stars (viz. Slettebak 
\cite{Sle56}, Conti \& Ebbets \cite{Con77}, Howarth et al. \cite{How97}); 
however, its behavior is still not well understood. To a first order we 
will consider an isotropic macroturbulence described by a Gaussian 
distribution of velocities. Some notes on other possible descriptions of 
the macroturbulent broadening are given in Sect. \ref{section4}.

Finally, in some cases, noise was added to the synthetic spectra by
using a Poissonian distribution of noise generated with {\sc randomn} 
in \idl.\\

A first test was performed to study how well does this method work when 
different lines from various elements present in the stellar spectrum of
OB stars are considered. Our main purpose here is to see whether the
Stark broadening affects the use of H and He lines as \vsini\ indicators.
This is particularly important when comparing metal and \ion{He}{} lines, 
as the latter have been sometimes proposed for \vsini\ measurements in 
early type stars using the \fwhm method, particularly when the metal
lines were faint and shallow, either because of the very early spectral type 
or the large rotational velocity.

\begin{figure}[t!]
\centering
\includegraphics[width=6.2cm,angle=90]
{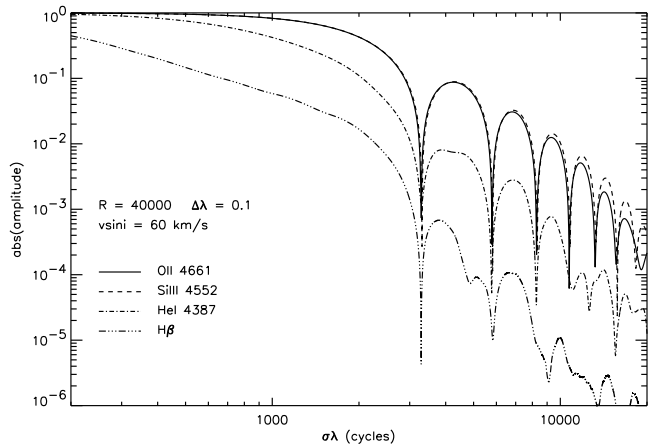}
\caption{Four different synthetic line profiles generated with \fastwind\ 
were convolved with a \vsini\ of 60 \kms. The different lines show the \ft\ 
of the lines on a $\sigma\lambda$ baseline. The first zero fits perfectly 
to its theoretical value for all the lines. The 
\ft\ can detect the \vsini\ feature even for the \ion{He}{i}\,$\lambda$4387 
and H$_{\beta}$ lines. Note that the \ft\ of the \ion{H}{} line reaches 
the value of 1 for $\sigma\lambda\,<$\,200}
\label{figure1}
\end{figure}

To graphically compare results from different lines, a $\sigma\lambda$ 
baseline was used, as the position of the first zero associated with the 
\vsini\ depends on $\lambda$ (this is equivalent to a $1/\rm v$ 
scale, see Eq. \ref{form1}). Fig. \ref{figure1} shows the result of 
this study. The Fourier technique has been applied to several \fastwind\ 
synthetic lines from different elements. As it can be seen, the agreement 
between the various lines is excellent (even when the lines 
\ion{He}{i}\,$\lambda$4387 or H$_{\beta}$ are used) which indicates that 
using the Fourier method even the strongly Stark broadened Balmer lines 
can be used, in principle, as \vsini\ diagnostics.
Note that the power of the Fourier transform of the \ion{He}{i} and H 
lines concentrates at low frequencies, diminishing more rapidly than for 
the other lines when larger frequencies are considered. This is an effect 
produced by the presence of non-rotational broadening mechanisms affecting 
the line profile (i.e. Stark broadening). Similarly to the case of lines 
affected by macroturbulence, it will have consequences in noisy spectra.\\
%
%
%
%

\begin{figure}[t!]
\centering
\includegraphics[width=6.8cm,angle=90]
{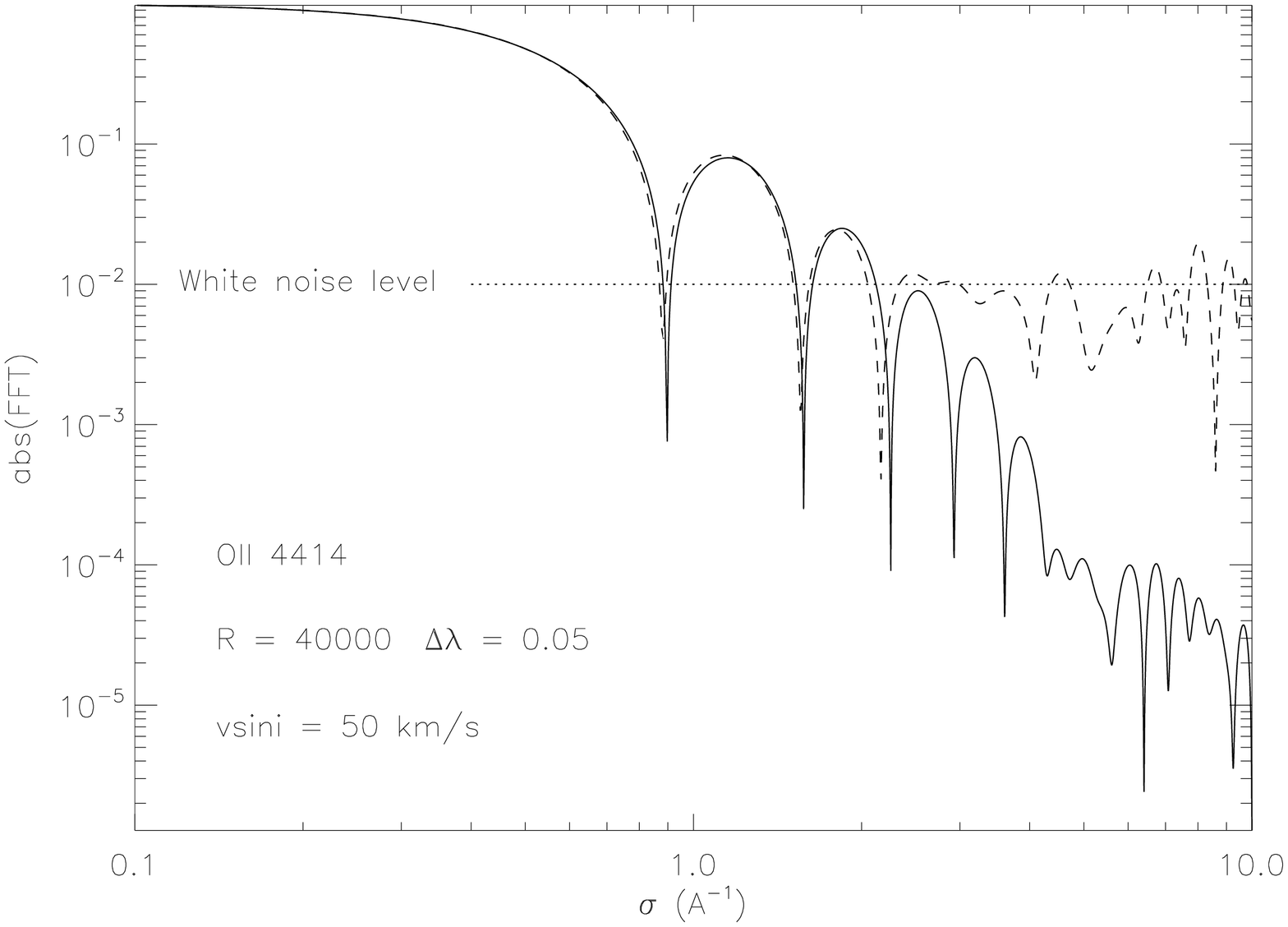}
\smallskip
\includegraphics[width=6.8cm,angle=90]
{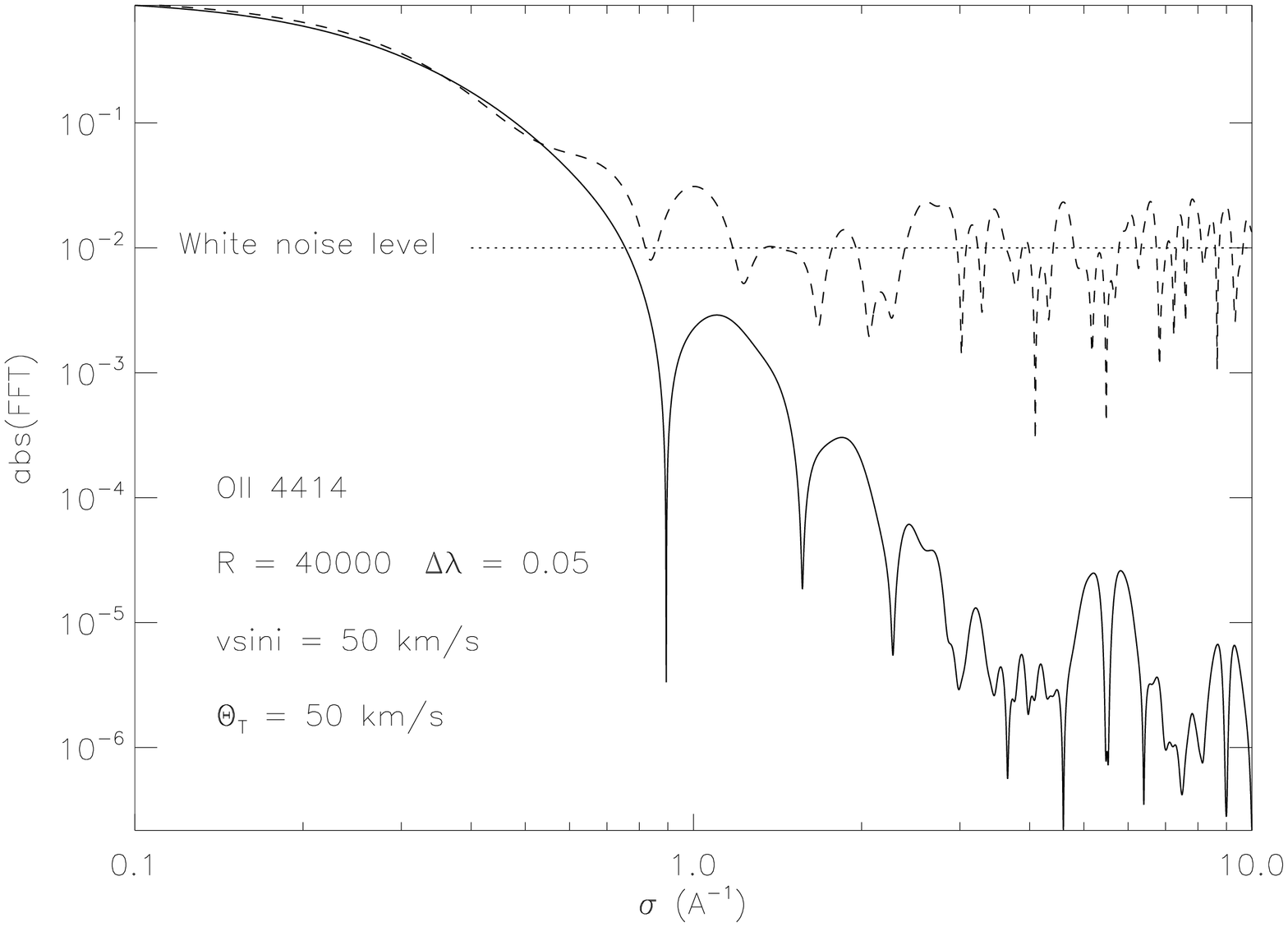}
\caption{Effect of the noise on the Fourier transform of a line profile. A
Poissonian distribution of noise generated with \idl, and corresponding to an
\snr\,=\,200, was added to a synthetic \ion{O}{ii} line, broadened to a 
\vsini\,=\,50 \kms\ (up), or to a combination of \vsini\,=\,50 \kms\ and 
\macro\,=\,50\,\kms\ (bottom). The figure compares the transforms of the line
with and without noise (dashed and solid lines, respectively).}
\label{figure2}
\end{figure}

How are these results affected by the noise present in real spectra? As
mentioned by Smith \& Gray (\cite{Smi76b}), random noise produces a "white 
noise" in the Fourier space (i.e. statistically constant with Fourier 
frequency). Therefore at higher frequencies, where the power of the signal 
transform drops, the "white noise" dominates and hides the characteristics 
of the signal transform. This effect is illustrated in Fig. \ref{figure2}. 
A strong synthetic \ion{O}{ii} line was convolved with two sets of rotational 
and macroturbulent broadenings: (\vsini, \macro)\,=\,(50, 0) and (50, 50) 
\kms. Additionally, a 
Poissonian distribution of noise, corresponding to a \snr\,=\,200, was 
added to both synthetic lines. Fig. \ref{figure2} compares the Fourier
transforms corresponding to the pure and noisy synthetic profiles. For
these cases, the white noise level is located at $\sim$\,10$^{-2}$. When
the profile is only affected by a rotational broadening the signal 
corresponding to the first zero is quite above the noise level. This first 
zero can be perfectly identified, and hence the \vsini\ can be accurately 
measured. The effect of the noise is more dramatic when the line is also 
broadened by macroturbulence; in this case, most of the power occurs at 
low frequencies in the Fourier domain and the noise level is above the 
signal transform even for the first rotational zero. Consequently, the 
unambiguous identification of the first zero in the Fourier transform 
becomes difficult or impossible, depending on the combination of \vsini, 
\macro\ and \snr.

A similar behavior will occur for the \ion{H}{}, \ion{He}{i} and \ion{He}{ii} 
lines. Although, in principle, it is possible to use these lines to derive 
the \vsini\ (as it has been show above), since these lines are broadened by 
mechanisms other than rotation, the signatures corresponding to this 
rotational broadening may be hidden below the noise level due to the 
lowering of the low frequencies power caused by Stark broadening.\\
%
%

In order to obtain an estimation of the feasibility of the \ft\ method 
when different quality (\snr) spectra are considered we constructed a grid 
of synthetic \linea{Si}{iii}{4552} lines convolved with various values of 
\vsini\ and \macro, and degraded to different \snr\ values. The \ft\ method 
was then used to determine the \vsini\ for the different cases (see results  
summarized in Table \ref{table1}). Although we do not expect the 
noise systematically affecting the position of the zeroes, this academic 
example should allow us to better understand the applicability limits of 
the \ft\ method in real spectra, affected by noise.

%
\begin{table}[!ht]
\begin{center}
\scriptsize{
\begin{tabular}{c | c c c | c c c}    
\hline
\hline
\noalign{\smallskip}
\vsini\ (\kms) &  & 60 & & & 30 & \\
\noalign{\smallskip}
\hline
\noalign{\smallskip}
\macro\ (\kms) & 0 & 40 & 80 & 0 & 40 & 80 \\
\noalign{\smallskip}
\hline
\noalign{\smallskip}
\snr\ = 500 & 60  & 60 & 64  & 30 & 32 & 35 \\
\snr\ = 300 & 60  & 60 & 67  & 26 & 35 & 47 \\
\snr\ = 100 & 62  & 66 & 75  & 32 & 46 & 47 \\
\snr\ = 50  & 59  & 65 & --- & 34 & 49 & 60 \\
\noalign{\smallskip}
\hline
\end{tabular}
\\
}
\normalsize 
\rm
\caption{\footnotesize Formal study of the effect of noise on the \vsini\ 
determination when a stellar line (\ion{Si}{iii}\,$\lambda$4452) is also 
affected by macroturbulence. We limit ourselves to low values because at 
higher \vsini\ the problem is much less significant. 
\label{table1}
}
\end{center}
\end{table}
%
In our formal study, we found that the effect of decreasing the \snr\ 
is to produce somewhat larger \vsini\ values when combined with macroturbulence. 
Although this effect is not 
very important when the macroturbulence value is small, deriving correct 
values even for \snr\,$\le$\,100, the effect may be more critical when the 
amount of macroturbulent broadening dominates over the rotational broadening. 
Note that for the case of \vsini\,=\,60 \kms\ and \macro\,=\,80 \kms\, a 
\snr\,$\ge$\,300 is required to be affected by an error of less than 
$\sim$\,10\,\% in the determination of the \vsini.
Finally, Table \ref{table1} shows how critical the effect of noise can be 
when the \vsini\ is low and the amount of macroturbulence is large (f.e. 
\vsini\,=\,30 \kms\ and \macro\,=\,80 \kms), a case that we may expect to 
find in supergiants.\\

\begin{figure}[ht!]
\centering
\includegraphics[width=6.3 cm,angle=90]{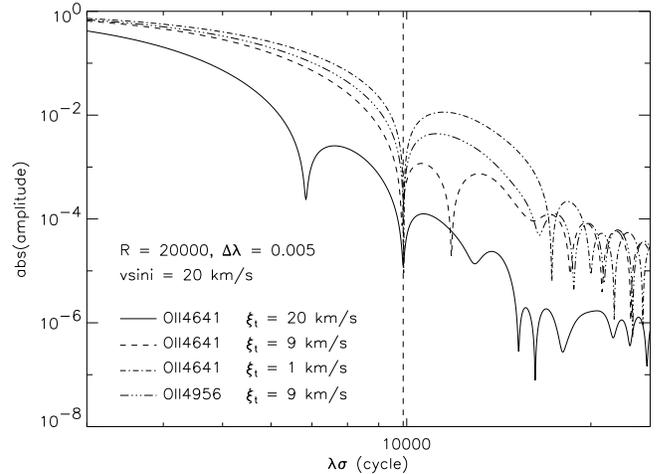}
\caption{Effect of the microturbulence over the determination of the \vsini\ 
         in the Fourier space.}
		 \label{figure3}
\end{figure}

The final test relates to the effect of microturbulence, which may introduce
additional zeroes as commented before. This is illustrated in Fig. 
\ref{figure3}. For this test a rotational velocity of 60 \kms\ has been 
adopted.
We see that the Fourier transform of the strong \ion{O}{ii}\,$\lambda$4641
line presents zeroes that are different when microturbulent velocities of
1, 9 and 20 \kms\ are introduced. With the only information from this line, 
it would be easy to get a wrong value for the projected rotational velocity
when using only the information from the zeroes (particularly if we only use
the first one). However, we also see in the figure that we can avoid this error
by using the information from weaker lines, like \ion{O}{ii}\,$\lambda$4956.
This line is not significantly affected by the microturbulence, and thus
only the zeroes due to rotation are simultaneously present in both transforms. 
%
%
\section{Applying the Fourier method to real OB stellar spectra}
\label{section3}
\begin{figure}[t!]
\centering
\includegraphics[width=6.8cm,angle=-90]{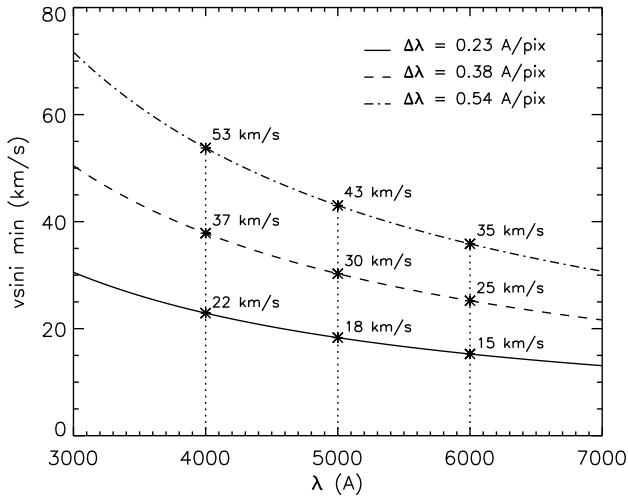}
\caption{Minimum \vsini\ that can be determined at different wavelengths for 
various spectral dispersions as a consequence of the combination of formula 
\ref{form1} with the Nyquist frequency}
\label{figure4}
\end{figure}

%
In the previous section it has been shown that the \ft\ technique allows
us a straightforward determination of projected rotational velocities.
When dealing with real spectra one has to remind two important 
properties of the FFT algorithm:
\begin{enumerate} 
\item {The sampling of the discrete Fourier transform ($\Delta\sigma$) 
inversely depends on the product of the number of points into which the 
line profile is sampled and the spectral dispersion: 
$\Delta\sigma$\,=\,(N$\Delta\lambda$)$^{-1}$.
Therefore, the better the sampling of the Fourier transform of the 
line profile, the more accurate the position of the zeroes associated 
to \vsini\ can be determined.} 
\item {The FFT of a sequence of $N$ real numbers is symmetric and hence 
the maximum $\sigma$ value that is computed is 
$\sigma_{\rm max}$\,=\,($N$/2)($N\Delta\lambda$)$^{-1}$\,=\,(2$\Delta\lambda$)$^{-1}$,
also known as Nyquist frequency.
From formula \ref{form1} and this computational property of the FFT it 
is found that the spectral dispersion ($\Delta\lambda$) of the stellar 
spectrum imposes a limit in the lowest \vsini\ that can be derived, given 
by 1.320\,c\,$\Delta\lambda$/$\lambda$. 
Fig. \ref{figure4} illustrates the minimum \vsini\ that can be measured for 
various characteristic $\Delta\lambda$ values.}
\end{enumerate}
%
%

Therefore, if the spectrum has enough resolution it will be possible to 
determine the projected rotational velocity of the star independently of 
any other broadening mechanism affecting the line profiles. The applicability 
of the \ft\ method in real spectra is mainly limited by the spectral 
dispersion (for low \vsini\ cases) and the \snr\ (mainly affecting 
those stellar lines broadened by a non-rotational broadening mechanism, i.e. 
macroturbulence and/or Stark broadening). \\

Having these ideas in mind, a sample of stellar spectra of OB stars, obtained 
during various observing runs, was restored from our archive to determine
their projected rotational velocities by means of the \ft\ method.
The derived values were compared with those resulting from the use
of the \fwhm\ method. This allowed us to test the validity of the \ft\ 
method by studying those cases in which stellar lines are mainly broadened 
by rotation, as well as the range of 
applicability of the \fwhm\ method. Additionally, the possible presence 
of macroturbulence could be studied.
%
\subsection{The \fwhm\ method}
\label{section31}
%
The \fwhm\ method requires the use of an unbroadened line template 
that is convolved with rotational profiles corresponding to various \vsini\ 
values. Several authors have considered different possibilities for the 
unbroadened profiles, from sharp-lined stars observed with the same 
configuration than the unknowns (viz. Slettebak \cite{Sle56}, who used 
10\,Lac; Penny \cite{Pen96}, who used HD\,34078; and Howarth et al. 
\cite{How97}, who used $\tau$\,Sco), to 
synthetic lines from stellar atmosphere code calculations (viz. 
Slettebak et al. \cite{Sle75}, Conti \& Ebbets \cite{Con77}),
or even Gaussian profiles with the same {\sc ew} than the observed line 
(viz. Herrero et al. \cite{Her92}). By determining the \fwhm\ from a 
Gaussian fit to the rotationally broadened line templates a \fwhm\,-\,\vsini\ 
calibration is then generated\footnote {Note that there is also the 
possibility of generating a \fwhm\,-\,\vsini\ calibration based in 
previous \vsini\ determinations (viz. Abt et al. \cite{Abt02}, Strom et 
al. \cite{Str05}).}. In this study we considered the same approach than 
Herrero et al. (\cite{Her92}).
%
\subsection{The observations}
\label{section32}
%
We have selected a sample of Galactic OB stars from observations carried
out with the Isaac Newton 2.5m and the William Herschel 4.2m telescopes 
at the Observatorio de El Roque de los Muchachos during various 
observational campaigns between 1989 and 2003. The Intermediate Dispersion 
Spectrograph ({\sc ids}) and the medium-resolution {\sc isis} spectrograph 
attached to these telescopes were used. The characteristics of the different 
observing runs are summarized in Table \ref{table2}.\\  
%
\begin{table*}[!t]
\begin{center}
\scriptsize{
\begin{tabular}{c c c c c c}    
\hline
\hline
\noalign{\smallskip}
Obs. C. & Date & I. config. & $\Delta\lambda$ (\AA/pix) & $R$ & $\lambda$ coverage \\
\noalign{\smallskip}
\hline
\noalign{\smallskip}
INT89  & Jul. 1989  & IDS235\,-\,H2400B & 0.38 & 7500  & 4000\,-\,5000 \AA \\
INT92  & Sept. 1991 & IDS235\,-\,H2400B & 0.23 & 9000  & 4000\,-\,5100 \AA \\
       &     ---    & IDS235\,-\,H1800V & 0.23 & 9000  & 6500\,-\,6700 \AA \\
WHTB   &            & ISIS\,-\,R1200B   & 0.23 & 9000  & 4000\,-\,4750 \AA \\
INT02  & Dec. 2002  & IDS235\,-\,H2400B & 0.23 & 9000  & 4050\,-\,5100 \AA \\
       &    ---     & IDS235\,-\,H1800V & 0.23 & 9000  & 6500\,-\,6700 \AA \\
WHT03  & Jun. 2003  & ISIS\,-\,R1200B   & 0.23 & 9000  & 3900\,-\,5100 \AA  \\
       &   ---      & ISIS\,-\,R1200B   & 0.23 & 9000  & 5500\,-\,5900 \AA \\
	   &   ---      & ISIS\,-\,R1200B   & 0.23 & 9000  & 6500\,-\,6700 \AA \\
CASPEC &            & CASPEC@\,ESO3.6   & 0.10 & 18000 & 4050\,-\,5000 \AA  \\
\noalign{\smallskip}
\hline
\\
\end{tabular}
}
\normalsize 
\rm
\caption{\footnotesize Observations used for the present work. First
column identifies the campaign by telescope and year; second column gives 
the observation date. The different instrument configurations used in each 
campaign, along with their spectral dispersion, approximate resolution, and 
observed spectral range, are specified in columns third to sixth, 
respectively.
\label{table2}
}
\end{center}
\end{table*}
%

The spectra were selected to have \snr\,$\ge$\,200 and spectral dispersion 
$\le$ 0.4 \AA/pix (which implies a minimum \vsini\ of $\sim$\,30\,-\,40 \kms, 
see Fig. \ref{figure4}). Only for late O-type and early B-type dwarfs 
some spectra were selected to have a \snr\ slightly below 200, as these 
targets are not expected to be affected by macroturbulence and hence the 
effect of noise is less critical when applying the \ft\ method to metal 
lines. The sample of stars ranges in spectral types from O3 to B2, including 
dwarfs, giants and supergiants (see Tables \ref{table3} to 
\ref{table9} and Table \ref{table10}).\\

The characteristics of the optical spectra of OB stars do not make it 
possible to use the same set of stellar lines to determine the projected 
rotational velocities for the whole sample of OB stars.
For example, while in late O-type and early B-type stars many metal lines 
from different elements are present in the stellar spectra, along with H 
and \ion{He}{i} lines, for the earliest O-type stars the spectrum is mainly 
dominated by H and \ion{He}{ii} lines, and only a few faint metal lines can 
be found.
Therefore we decided to perform the study separately following the 
classification presented in Sects \ref{section33} to \ref{section36}.
%
\subsection{Early B and late O dwarfs: an easy task}
\label{section33}
\begin{figure*}[t!]
\centering
\includegraphics[width=6.8cm,angle=90]
{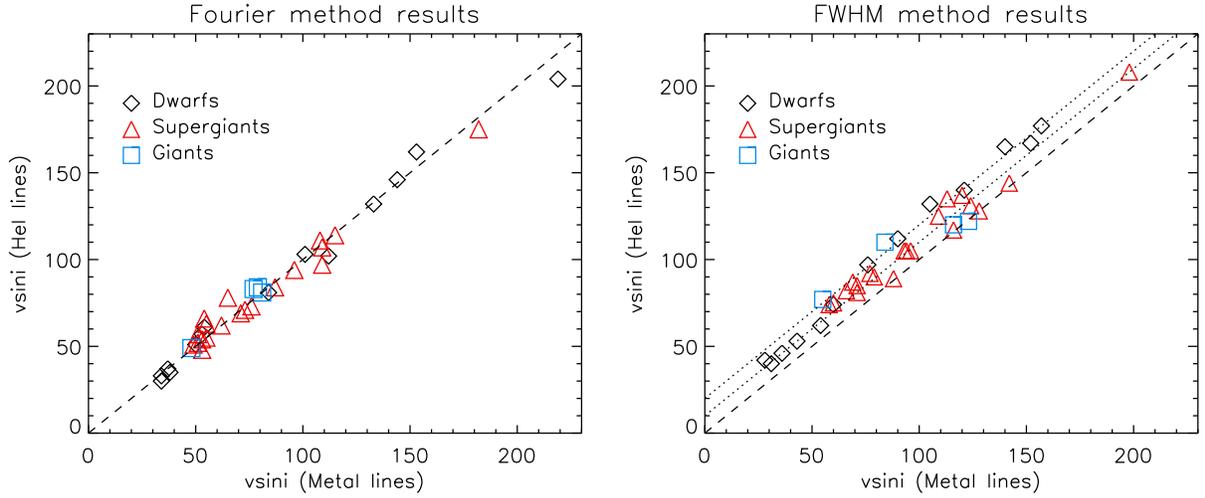}
\caption{Comparison of \vsini\ results derived from metal and \ion{He}{i} 
lines through the \ft\ and \fwhm\ methods. Only those cases in which \vsini\ 
could be derived through metal lines are presented. Dotted lines in the 
\fwhm\ diagram represents a difference between both determinations of 10 
and 20 \kms.}
\label{figure5}
\end{figure*}

\begin{figure}[t!]
\centering
\includegraphics[width=6.9cm,angle=90]
{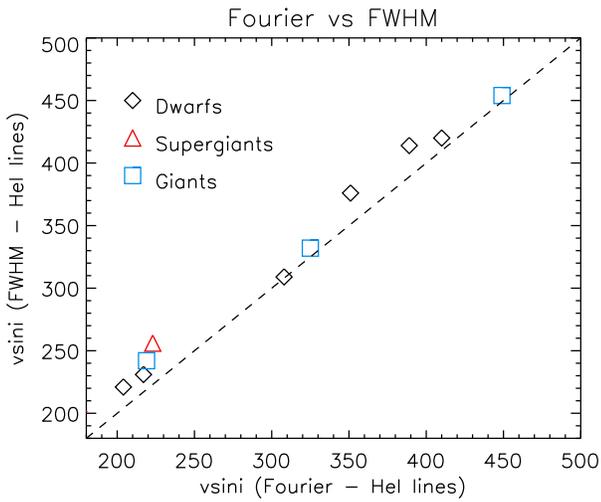}
\caption{Comparison of \vsini\ results derived from \ion{He}{i} lines 
through the \ft\ and  \fwhm\ methods for \vsini\,$\ge$\,180 \kms. In this
range of projected rotational velocities, no metal lines are accessible 
for the \vsini\ determination.}
\label{figure6}
\end{figure}

\begin{figure}[t!]
\centering
\includegraphics[width=6.7cm,angle=90]
{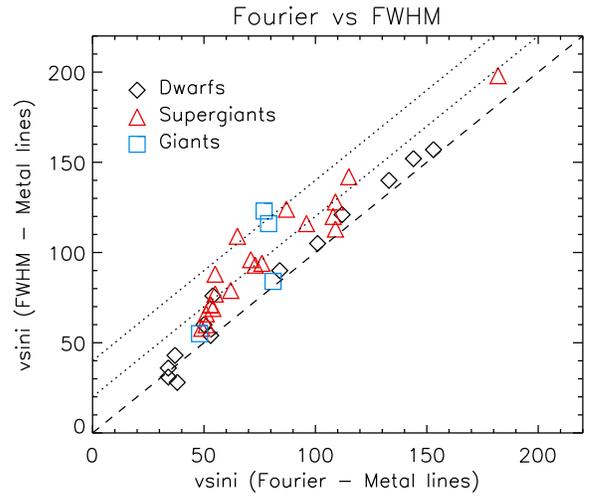}
\caption{Comparison of \vsini\ results derived from metal
lines through the \ft\ and \fwhm\ methods. Dotted lines represent
differences between both determinations of 20 and 40 \kms.}
\label{figure7}
\end{figure}

%
Early B and late O dwarfs have many strong metal lines, along with some 
\ion{He}{i} lines, useful to infer the stellar \vsini. 
We do not expect important broadening mechanisms other than rotation 
affecting the metal lines; however, the He lines are additionally broadened 
by the Stark broadening. Tables \ref{table3} and \ref{table4} show the 
\vsini\ derived from several \ion{He}{i}, \ion{Si}{iii}, \ion{C}{ii}, 
\ion{N}{ii}, and \ion{O}{ii} lines present in the spectra of dwarf stars 
with spectral types between B1 and O9. Results obtained through both the 
\ft\ and \fwhm\ methods are presented.
A comparison of \vsini\ results is shown in Figures \ref{figure5} to
\ref{figure7} (dwarfs are presented as black diamonds).\\
\newline
For dwarfs, three main conclusions arise from this comparison:
\begin{itemize}
\item{\ion{He}{i} lines give results consistent with metal lines when 
analyzed with the \ft\ method (Fig. \ref{figure5}, left), but tend to 
produce larger values for the projected rotational velocities when using 
the \fwhm\ method (10\,-\,20\,\%, see Figure \ref{figure5}, right). 
For very large rotational velocities, when no metals can be used, 
\ion{He}{i} lines give slightly larger \vsini\ values when using the \fwhm\ 
method than when using the \ft, although the effect is smaller than for 
lower \vsini\ (see Fig. \ref{figure6}).}
\item{From Fig. \ref{figure7} it can be seen that for O9\,-\,B0 dwarfs 
both methods, the \ft\ and \fwhm, give similar results when metal lines 
are used.}
\item{The shape of the Fourier transform of the metal lines is very close 
to that resulting from a pure rotational profile, confirming the absence 
of important broadening mechanisms other than rotation. This fact allows 
an accurate determination of the \vsini\ of late-O and early-B dwarfs 
even for \snr\,$\sim$\,120.}
\end{itemize}
%
%
%
\begin{table*}[!t]
\begin{center}
\scriptsize{
\begin{tabular}{c c c c c c c c c}    
\hline
\hline
\noalign{\smallskip}
Star & $\omega^1$\,Sco & HD\,217657 & HD\,37020 & HD\,37023 & HD\,37042 & $\tau$\,Sco & CygOB2-145 & HD214680 \\
SpT & B1V & B0.5V & B0.5V & B0.5V & B0.5V & B0.2V & O9.5V & O9V \\
Obs. run & WHT03 & INT92 & INT02 & INT02 & INT02 & CASPEC & WHT03 & WHT03 \\
\snr & 350 & 120 & 220 & 290 & 350 & 250& 120 & \\
\noalign{\smallskip}
\hline
\noalign{\smallskip}
\linea{He}{i}{4713}   & 102 (115) & 31 (38) & 53 (58) & 50 (69) & 31 (46) & 14 (  ) & 35 (35) & 36 (48) \\
\linea{He}{i}{4922}   & 104 (178) &  b      &  b      &  b      &    b    &    b    &   b     &  b \\
\linea{He}{i}{5015}   & 101 (114) & 29 (35) & 60 (63) & 48 (77) & 30 (45) & 15 (  ) & 28 (34) & 35 (50) \\
\linea{He}{i}{5047}   & 102 (125) & 32 (36) & 52 (55) & 52 (70) & 32 (39) & 18 (  ) & 28 (30) & 33 (46) \\
\linea{He}{i}{5875}   & 104 (119) &  -      &  -      &  -      &  -      &  -      & 48 (59)  & 47 (64 ) \\
\linea{He}{i}{6678}   & 105 (120) & 29 (49) & 62 (70) & 55 (80) & 35 (53) &  -      & 34 (50) & 36 (58 ) \\
\noalign{\smallskip}
\linea{Si}{iii}{4552} & 101 (110) & 36 (21) & 56 (56) & 52 (72) & 32 (36) & 10 (13) & 40 (27) & 34 (40) \\
\linea{Si}{iii}{4567} & 104 (110) & 31 (33) & 49 (58) & 50 (69) & 33 (37) & 12 (14) & 41 (20) & 38 (46) \\
\linea{Si}{iii}{4574} & 103 (114) & 39 (35) & 53 (54) & 54 (70) & 31 (33) & 11 (13) & 39 (24) & 32 (43) \\
\noalign{\smallskip}
\linea{C}{ii}{4267}   &  97 (100) & 41 (36) & 49 (53) & 54 (75) & 32 (37 ) & 21 (22) & 36 (38 )  & 40 (44) \\
\noalign{\smallskip}
\linea{N}{ii}{3995}   & 103 (102) & 44 (28 )    &  -    &  -      &  -    &    -     &  -    & 45 (49 ) \\
\linea{N}{ii}{4253}   & 96 (99)   & 43 (44) & 56 (60) & 55 (66) & 36 (44) &    b    & 43 (36) & 38 (50) \\
\linea{N}{ii}{4601}   & 104 (110) &   r     & 53 (50) & 54 (62) &  b?     &    b    &  d       & 35 (42) \\
\noalign{\smallskip}
\linea{O}{ii}{4317}   &  b        & 36 (35) & 56 (54) & 51 (55) & 37 (38) & 14 (12) & 44 (31) &  d \\
\linea{O}{ii}{4319}   &  b        & 36 (36) & 58 (54) & 47 (53) & 38 (35) & 12 (14) & 39 (25) &  39 (42) \\
\linea{O}{ii}{4414}   &  b        & 32 (25) & 49 (52) & 48 (50) & 40 (36) & 14 (13) & 35 (29 )  &  d \\
\linea{O}{ii}{4416}   &  b        & 30 (29) & 52 (49) & 47 (61) & 33 (36) & 10 (14) &  r      &  d \\
\linea{O}{ii}{4590}   & 108 (109) & 32 (28) & 54 (55) & 48 (63) & 32 (35) & 12 (13) & 40 (48) &  b \\
\linea{O}{ii}{4595}   & 97 (105)  & 35 (36) & 56 (54) & 48 (52) & 35 (37) & 12 (16) & 42 (24) &  39 (44) \\
\linea{O}{ii}{4661}   & 102 (99)  & 28 (24) & 55 (51) & 51 (55) & 30 (33) & 14 (13) & 33 (18) & 37 (37) \\
\linea{O}{ii}{4699}   &  b        & 29 (35) & 51 (56) & 48 (54) & 35 (38) &   b     & 33 (34 )  &  37 (44) \\
\linea{O}{ii}{4705}   &  b        & 30 (21) & 48 (52) & 44 (52) & 33 (33) & 14 (14) & 35 (28 )  &  34 (37) \\
\linea{O}{ii}{4906}   & 100 (100) &  d      & 57 (52) & 46 (51) & 40 (37) & 16 (12) &  d      &  d \\
\linea{O}{ii}{4943}   &  b        & 29 (26) & 53 (57) &     b   & 34 (30) & 12 (13) & 29 (21) &  d \\
\noalign{\smallskip}
\hline
\noalign{\smallskip}
\noalign{\smallskip}
\hline
\noalign{\smallskip}
\vsini\ (FT-He) & \solu{103}{2} & \solu{30}{2} & \solu{57}{5} & \solu{51}{3} & \solu{33}{4} & \solu{16}{2} & \solu{35}{8}  & \solu{37}{6} \\
\vsini\ (FW-He) & \solu{132}{24}& \solu{40}{6} & \solu{62}{7} & \solu{74}{5} & \solu{46}{6} & \solu{}{}    & \solu{42}{12} & \solu{53}{8} \\
\vsini\ (FT-M)  & \solu{101}{4} & \solu{34}{5} & \solu{53}{3} & \solu{50}{3} & \solu{34}{3} & \solu{13}{3} & \solu{38}{4}  & \solu{37}{3} \\
\vsini\ (FW-M)  & \solu{105}{5} & \solu{31}{6} & \solu{54}{3} & \solu{60}{8} & \solu{36}{4} & \solu{14}{3} & \solu{28}{8}  & \solu{43}{4} \\
\noalign{\smallskip}
\hline
\noalign{\smallskip}
\end{tabular}
}
\normalsize 
\rm
\caption{\footnotesize Projected rotational velocities (\vsini\ in \kms)
derived from several 
\ion{He}{i}, \ion{Si}{iii}, \ion{C}{ii}, \ion{N}{ii}, and \ion{O}{ii} 
lines present in the spectra of dwarf stars with spectral types between B1 
and O9. Values in brackets represent \fwhm\ results. Mean values for 
\ion{He}{i} and metal lines are also presented, with errors showing the 
standard deviation of line values. (b) blended line, (r) noise is important, 
(d) faint line, (-) line not present in the observed spectrum (either 
because that line does not appear for this SpT, or because it is out of 
the observed spectral range).
\label{table3}
}
\end{center}
\end{table*}
%
%
\begin{table*}[!t]
\begin{center}
\scriptsize{
\begin{tabular}{c c c c c c c}    
\hline
\noalign{\smallskip}
Star & HD\,37061 & HD\,228199 & CygOB2-23 & BD-134928 & HD\,149757 & HD\,37041 \\
SpT & B1V & B0.5V & O9.5V & O9.5Vn & O9.5Vnn & O9V \\
Obs. run & INT02 & INT89 & WHT03 & WHT03 & WHT03 & INT02 \\
\snr & 250 & 100 & 190 & 230 & 500 &  230 \\
\noalign{\smallskip}
\hline
\noalign{\smallskip}
\linea{He}{i}{4387}   & 213 (235) & 109 (146) & 158 (185) &  404 (433) & 360 (440) & 132 (180) \\
\linea{He}{i}{4471}   & 213 (270) &  91 ()    & 159 (212) &  b         & 377 (433) & 141 (200) \\
\linea{He}{i}{4713}   & 220 (224) &  99 (114) & 167 (160) &  d         & b         & 127 (145) \\
\linea{He}{i}{4922}   & 218 (250) & 107 (159) & 177 ()    &  405 (416) & 389 (402) & 125 (170) \\
\linea{He}{i}{5015}   & 216 (206) &  -        & 153 (162) &  401 (415) & 410 (411) & 135 (138) \\
\linea{He}{i}{5047}   &     (212) &  -        &  d        &  d         & 395 (420) & 130 (166) \\
\linea{He}{i}{5875}   &     -     &  -        & 154 (166) &  429 (421) & 385 (387) &  -  \\
\linea{He}{i}{6678}   & 220 (219) &  -        & 164 (176) &  b         & 404 (407) & 133 (158) \\
\noalign{\smallskip}
\linea{Si}{iii}{4552} &  d        & 114 (116) &  d          &  d     &  d   &  d  \\
\linea{Si}{iii}{4567} &  d        & 118 (127) &  d          &  d     &  d   &  d  \\
\linea{Si}{iii}{4574} &  d        & 103 (120) &  d          &  d     &  d   &  d  \\
\linea{Si}{iv}{4089}  &  d        &    b      &  153 (157)  &  b     &  b   &  133 (140) \\
\noalign{\smallskip}
\hline
\noalign{\smallskip}
\noalign{\smallskip}
\hline
\noalign{\smallskip}
\vsini\ (FT-He) & \solu{217}{3}  & \solu{102}{8}  & \solu{162}{8}  & \solu{410}{13} & \solu{389}{17} & \solu{132}{5} \\
\vsini\ (FW-He) & \solu{231}{23} & \solu{140}{23} & \solu{177}{20} & \solu{420}{8}  & \solu{414}{18} & \solu{165}{21} \\
\vsini\ (FT-M)  & -              & \solu{112}{8}  &       153      & -              &    -           & 133 \\
\vsini\ (FW-M)  & -              & \solu{121}{6}  &       157         & -              &    -           & 140 \\
\noalign{\smallskip}
\hline
\noalign{\smallskip}
\end{tabular}
}
\normalsize 
\rm
\caption{\footnotesize Same as in Table \ref{table3} but for stars with
\vsini\ values above 110 \kms. For these cases, the high \vsini\ makes metal
lines to be faint and blended, and hence not suitable to measure projected
rotational velocities. Except for a few metal lines, only \ion{He}{i} lines
can be used.
\label{table4}
}
\end{center}
\end{table*}
%

There are two cases (CygOB2-145, and $\tau$\,Sco) in which the measurement 
of \vsini\ is limited by the spectral resolution. The \vsini\ value resulting 
from the \ft\ analysis is very close to the limit imposed by the Nyquist 
frequency (i.e. 1.320\,$c\Delta\lambda$/$\lambda$). In addition, the \vsini\ 
derived from the \fwhm\ method is somewhat below the \ft\ value. A larger 
spectral dispersion ($\Delta\lambda$) is needed for these stars to correctly 
derive the \vsini\ through the \ft\ method. 
Moreover, even with better resolution spectra, the broadening produced by 
microturbulence may be comparable to that one corresponding to the rotational 
velocity, and hence zeroes associated with microturbulence may be found at 
frequencies somewhat smaller than $\sigma_1$ in the Fourier space. These stars 
may be good examples to study the effects illustrated in Fig. \ref{figure3}.\\

Although we are not going to treat it here in detail, we would like to mention
that some studies have shown that in the cases of large \vsini, the effect
of differential rotation may alter the position of zeros in the Fourier
transform. We refer the reader to Reiners \& Royer (\cite{Rei04}) for a
detailed study of this effects in the case of A-type stars. However, note that
Reiners \& Royer only find evidence of differential rotation in 4 out of
78 stars of spectral types later than ours, and therefore we don't expect 
this effect to be important in our sample.
%
%
\subsection{Early B and late O supergiants: the presence of extra broadening}
\label{section34}
\begin{figure}[h!]
\centering
\includegraphics[width=6.8cm,angle=-90]
{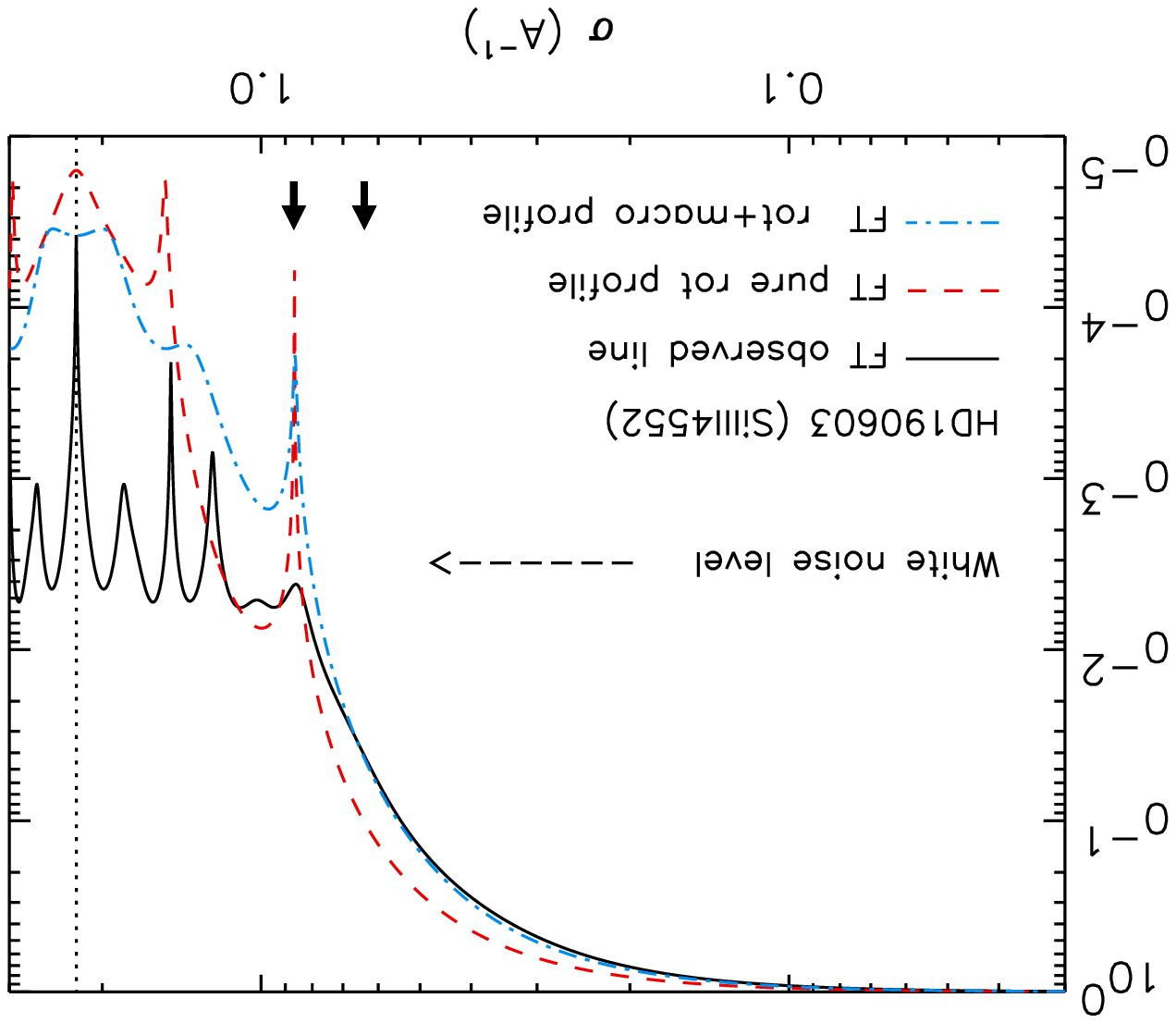}
\caption{Fourier transform of the \linea{Si}{III}{4552} line in the 
{\sc isis}@{\sc wht} spectrum of HD190603 (B1.5Ia$^+$). The \ft\
of the observed line (solid black) is compared with two synthetic profiles:
(dashed red) pure rotational profile with \vsini\,=\,50 \kms; (dashed-dotted 
blue) rotational\,+\,macroturbulent profile with (\vsini, \macro)\,=\,(50, 
45) \kms. Black arrows show the position of the zeroes associated with  
\vsini\ values of 50 and 68 \kms, corresponding to the values derived through 
the \ft\ and \fwhm\ methods, respectively.}
\label{figure8}
\end{figure}

%
The logical extension of the previous analyses points towards early B and
late O supergiants since these stars also have many narrow metal lines. 
Results for a sample of supergiants with spectral types ranging from 
O9\,I to B2\,I are presented in Table \ref{table5}, and in Figures 
\ref{figure5} to \ref{figure7}, the results from both the \ft\ and \fwhm\ 
methods are again compared (supergiants are shown as red triangles). \\
\newline
In this case of supergiants we find:
\begin{itemize}
\item {The \ft\ method gives again consistent results for both metal and 
\ion{He}{i} lines (Fig. \ref{figure5}).}
\item{For metal lines, the \vsini\ derived through the \fwhm\ method is 
systematically larger than that resulting from the \ft\ method (Fig. 
\ref{figure7}). This result confirms the presence of an extra broadening 
mechanism in early B-type supergiants, as shown by Dufton et al. 
(\cite{Duf06}), and extends the results towards late O-type supergiants.}
\item{An additional indication of the presence of an extra broadening is 
clearly shown by the shape of the Fourier transform of metal lines (see
e.g. Fig. \ref{figure8}). In this case, the power of the Fourier 
transform concentrates at higher frequencies when compared to the Fourier 
transform of a pure rotational line profile with the same EW.}
\end{itemize} 

In Sect. \ref{section21} it was shown that the effect of the white noise in the 
Fourier space can be important when an extra-broadening is present. If the \snr\ 
of the spectrum is not good enough, the zeros associated to the \vsini\ 
may be hidden below the white noise, and may be mistaken with artificial 
zeros associated with this white noise (see Fig. \ref{figure2}). A low \snr\
may result in larger \vsini\ derived values. 

In Fig. \ref{figure8}, the line \linea{Si}{iii}{4552} in the 
{\sc isis}@{\sc wht} spectrum of HD190603 (B1.5Ia$^+$) is used as
illustrative example of the application of the \ft\ method to an early B-type
Supergiant. The spectrum has a \snr\,$\sim$\,570 (see Table \ref{table5}).
The position of the first zero in the \ft\ of the observed line corresponds 
to a \vsini\,=\,50 \kms. The shape of the \ft\ of a pure rotational 
(+ instrumental!) profile corresponding to that value does not fit correctly 
the observed case for low frequencies. An extra macroturbulent broadening, 
\macro, is needed to produce a satisfactory fit in the Fourier space. Note 
that in this case the power of the \ft\ is below the white noise level of 
the observed line, hence indicating that the \ft\ of the observed profile 
may be affected by the white noise, even for those frequencies where the 
first zero associated with the rotation is located. The actual \vsini\ value 
may be somewhat smaller (and consequently, \macro\ larger).
%
%
\begin{table*}[!ht]
\begin{center}
\scriptsize{
\begin{tabular}{c c c c c c c c}    
\noalign{\smallskip}
\hline
\hline
\noalign{\smallskip}
Target   &  HD\,210809 & CygOB2-10 & HD\,209975 & HD\,167264 & HD\,37128 & HD\,38771 & HD\,2905 \\
SpT      &  O9I        & O9.5I     & O9.5Iab    &  B0Ia      & B0Ia      & B0.5Ia    & BC0.7Ia  \\
Obs. run &  INT89      & WHT03      & INT89      & WHTB       & WHTB      & WHTB      & WHTB    \\
\snr     &  200        & 250        & 200        & 400        & 330       & 400       & 300     \\
\noalign{\smallskip}
\hline
\noalign{\smallskip}
\ion{C}{ii}\,4267   &          &     r    &     -    & 73 (88) & 75 (94)  & 67 (80) & 69 (78) \\ 
\noalign{\smallskip}
\ion{N}{ii}\,3995   &           &     d    &    -     & 67 (72) & 64? (89) & 58 (80) & 58 (66) \\ 
\ion{N}{ii}\,4254   &           &     r  &    -     & 75 (94) & 69 (92)  & 57 (89) & 64 (83) \\ 
\ion{N}{ii}\,4601   &           &     -    &    -     & -       & 81 (91)  & 59 (89) & 72 (81) \\ 
\ion{N}{ii}\,4614   &           &     -    &    -     & -       & -        & -       & -        \\ 
\ion{N}{ii}\,4621   &           &     -    &    -     & -       & -        & -       & -        \\ 
\ion{N}{ii}\,4630   &           &  79 (80) &  69 (71) & 65 (89) & -        & 55 (94) & 62 (78)  \\  
\noalign{\smallskip}
\ion{Si}{iii}\,4552 &           & 63 (93)  &    r     & 73 (100) & 83 (98)  & 53 (98) & 64 (87)  \\  
\ion{Si}{iii}\,4567 &           & 76 (101) &    r     & 76 (100) & 83 (98)  & 52 (96) & 55 (87)  \\  
\ion{Si}{iii}\,4574 &           & 81 (99)  &    r     & 77 (103) & r (77)   & r (94)  & 52 (81)  \\ 
\ion{Si}{iv}\,4089  & 105 (125) & 78 (116) & 59 (110) & 72 (114) & 79 (112) & 73 (99) & r (86)   \\  
\ion{Si}{iv}\,4116  & 110 (115) & 72 (93)  & 70 (108) & (101)    & 65 (98)  & 51 (80) & 67 (70)  \\  
\noalign{\smallskip}
\ion{O}{ii}\,4590   &           & 76 (106) &    -     & 75 (92)  & 81 (94)  & 51 (86) & 66 (81)  \\ 
\ion{O}{ii}\,4595   &           &     r    &    -     & 52 (82)  & 81 (94)  & 52 (85) & 58 (82)  \\ 
\ion{O}{ii}\,4661   &           & 61 (76)  &    d     & 73 (82)  & 74 (96)  & 53 (86) & 55 (81)  \\ 
\noalign{\smallskip}
\ion{He}{i}\,4143  &            &      r   & 70 (115) & 68 (100) & 78 (94)  & 68 (96) & 69 (83)  \\ 
\ion{He}{i}\,4387  & 108 (140)  & 70 (102)  & 80 (132) & 75 (107) & 79 (104) & 71 (99) & 65 (99)  \\ 
\ion{He}{i}\,4471  & 113 (150)  & 76 (113)  &          & 73 (120) & 74 (127) &  b      &      b   \\ 
\ion{He}{i}\,4713  & 111 (127)  & 61 (93)   & 75 (117) & 65 (88)  & 67 (85)  & 55 (78) & 59 (81)  \\ 
\ion{He}{i}\,4922  & 110 (130)  & 76 (114)  & -        & -        & -        & -       & -        \\ 
\ion{He}{i}\,5015  &     -      & 60 (98)   & -        & -        & -        & -       & -        \\ 
\ion{He}{i}\,5047  &     -      & 61 (85)   & -        & -        & -        & -       & -        \\ 
\ion{He}{i}\,5875  &     -      &    r      & -        & -        & -        & -       & -        \\ 
\ion{He}{i}\,6678  &     -      &    r      & -        & -        & -        & -       & -        \\ 
\noalign{\smallskip}
\hline
\noalign{\smallskip}
\vsini\ (FT-He) & \solu{111}{2}  & \solu{67}{8}  & \solu{78}{4}   & \solu{71}{5}   & \solu{73}{6}   & \solu{63}{11} & \solu{62}{3}  \\
\vsini\ (FW-He) & \solu{137}{10} & \solu{100}{11} & \solu{125}{11} & \solu{105}{16} & \solu{105}{21} & \solu{89}{15} & \solu{90}{13} \\
\vsini\ (FT-M)  & \solu{108}{4}  & \solu{73}{7}   & \solu{66}{6}   & \solu{73}{4}   & \solu{76}{7}   & \solu{55}{5}  & \solu{62}{6}  \\
\vsini\ (FW-M)  & \solu{120}{7}  & \solu{95}{13}  & \solu{103}{10}  & \solu{93}{11}  & \solu{94}{8}   & \solu{88}{7}  & \solu{79}{6}  \\
\noalign{\smallskip}
\hline
\noalign{\smallskip}

\noalign{\smallskip}
\hline
\hline
\noalign{\smallskip}
Target   &  CygOB2-2 & HD\,13854 & HD\,190603 & HD\,14956 & HD\,14143 & HD\,14818 & HD\,194279  \\
SpT      &  B1I      & B1Iab     & B1.5Ia+    & B1.5Ia    & B2Ia      & B2Ia      & B2Ia        \\
Obs. run &  WHT03    & WHTB      & WHTB       & WHTB      & WHTB      & WHTB      & WHTB        \\
\snr     &           & 250       & 570        & 450       & 450       & 450       & 450         \\
\noalign{\smallskip}
\hline
\noalign{\smallskip}
\ion{C}{ii}\,4267   & 43 (50)  & 54 (78)  & 49 (54) &  (69)    & 42 (66)  & 68 (76) & 55 (69)  \\ 
\noalign{\smallskip}
\ion{N}{ii}\,3995   &          & 48 (79)  & 51 (64) & 46? (72) & 63? (70) & 48 (69) & 54 (74)  \\ 
\ion{N}{ii}\,4254   & 59 (64)  & 57 (77)  & 52 (62) & 56 (67)  & 51 (64)  & 49 (63) & 55 (73)  \\ 
\ion{N}{ii}\,4601   & 68 (68)  & 61? (81) & 53 (60) & 55 (77)  & 54 (66)  & 52 (73) & 52 (74)  \\ 
\ion{N}{ii}\,4614   &          & 60 (68)  & 44 (53) & 53? (68) & 56 (59) & 50 (52)  & 53 (67)  \\ 
\ion{N}{ii}\,4621   &          & b (73)   & 49 (54) & 52 (69)  & 52 (67) & 50 (68)  & 52 (66)  \\ 
\ion{N}{ii}\,4630   &          & 52 (84)  & 52 (63) & 54? (76) & 53 (73) & 54 (70)  & 59 (74)  \\  
\noalign{\smallskip}
\ion{Si}{iii}\,4552 &  44 (58) & r (85)   & 50 (68) & 55 (76)  & 48 (70) & 47 (74)  & 49 (75)  \\  
\ion{Si}{iii}\,4567 &  48 (61) & 45 (83)  & 44 (68) & 55 (75)  & 54 (72) & 54 (74)  & 52 (71)  \\  
\ion{Si}{iii}\,4574 &  45 (55) & 54 (77)  & 52 (63) & 61 (68)  & (66)    & 60 (68)  & 52 (68)  \\ 
\ion{Si}{iv}\,4089  &          & r (73)   &  b      &   d      &  d      &  d       &  d   \\  
\ion{Si}{iv}\,4116  &  58 (56) & 65 (73)  & 49 (59) &   d      &  d      &  d       &  d  \\  
\noalign{\smallskip}
\ion{O}{ii}\,4590   &  42 (67) & r (74)   & 57 (59) & 61 (69)  & 44 (66) & 57 (64)  & 57 (67)  \\ 
\ion{O}{ii}\,4595   &  44 (63) & r (67)   & 55 (56) & 43 (69)  & 46 (54) & 53 (67)  & 56 (71)  \\ 
\ion{O}{ii}\,4661   &  47 (41) & r (76)   & 52 (58) & 48 (69)  & 47 (59) & 54 (68)  & 48 (67)  \\ 
\noalign{\smallskip}
\ion{He}{i}\,4143  &     b    & 62(86)    & 54 (67) & 48 (80) & 53 (71) & 57 (81)  & 56 (76)  \\ 
\ion{He}{i}\,4387  &  50 (92) & 48(90)    & 46 (74) & 48 (87) & 53 (81) & 61 (85)  & 59 (82)  \\ 
\ion{He}{i}\,4471  &     b    & 62(110)   & 59 (88) & 73 (98) & 67 (98) & 74 (103) & 66 (95)  \\ 
\ion{He}{i}\,4713  &  47 (63) & 56 (77)   & 50 (62) & 47 (69) & 52 (66) & 64 (72)  & 48 (65)  \\ 
\ion{He}{i}\,4922  &    b     & -         & -        & -        & -        & -       & -        \\ 
\ion{He}{i}\,5015  &  49 (67) & -         & -        & -        & -        & -       & -        \\ 
\ion{He}{i}\,5047  &  49 (58) & -         & -        & -        & -        & -       & -        \\ 
\ion{He}{i}\,5875  &  59 (82) & -         & -        & -        & -        & -       & -        \\ 
\ion{He}{i}\,6678  &  50 (79) & -         & -        & -        & -        & -       & -        \\ 
\noalign{\smallskip}
\hline
\noalign{\smallskip}
\vsini\ (FT-He) & \solu{51}{5}  & \solu{55}{7}  & \solu{52}{7}  & \solu{48}{1}  & \solu{53}{1}  & \solu{63}{2}  & \solu{54}{8}  \\
\vsini\ (FW-He) & \solu{74}{13} & \solu{92}{17} & \solu{75}{13} & \solu{85}{15} & \solu{82}{16} & \solu{87}{16} & \solu{87}{15}  \\
\vsini\ (FT-M)  & \solu{49}{9}  & \solu{55}{6}  & \solu{51}{4}  & \solu{53}{5}  & \solu{51}{6}  & \solu{54}{6}  & \solu{53}{3}  \\
\vsini\ (FW-M)  & \solu{58}{8}  & \solu{77}{6}  & \solu{60}{5}  & \solu{71}{4}  & \solu{66}{6}  & \solu{69}{6}  & \solu{71}{3} \\
\noalign{\smallskip}
\hline
\noalign{\smallskip}
\end{tabular}
}
\normalsize 
\rm
\caption{\footnotesize Same as Table \ref{table3} but for Early-B and late-O supergiants
\label{table5}
}
\end{center}
\end{table*}
%
\begin{table*}[!t]
\begin{center}
\tiny{
\begin{tabular}{c c c c c c c c c c}    
\hline
\hline
\noalign{\smallskip}
Star     & HD\,12993 & CygOB2-16 & HD\,46056 & HD\,13268 & HD\,46966 & HD\,236894 & HD\,168137 \\
SpT      & O6.5V     & O7.5V     & O8Vn      & ON8V      & O8V       & O8V        & O8.5V      \\
Obs. run & PerOB1    & WHT03     & CAHA      & INT89     & INT89     & PerOB1     & WHT03      \\
\snr     &           &           &           &           &           &            &            \\
\noalign{\smallskip}
\hline
\noalign{\smallskip}
\linea{He}{i}{4387}  &  77 (123) & 209 (220) & 345 (352) &  315 (307) & 70 (100) & 148 (170) & 80 (106) \\
\linea{He}{i}{4471}  &  80 (90)  & 214 (230) & 350 (404) &  310 (320) & 54 (98)  & 142 (176) & 59 (110) \\
\linea{He}{i}{4713}  &  87 (122) & 203 (216) &           &            & 57 (72)  & 148 (156) & 48 (73)  \\
\linea{He}{i}{4922}  &   -       & 191 (217) & 359 (371) &  298 (301) & 54 (80)  &    -      & 58 (100) \\
\linea{He}{i}{5875}  &   -       & 201 (224) &     -     &     -      &          &    -      & 53 ()    \\
\noalign{\smallskip}
\linea{Si}{iv}{4089} & 84 (90)   &  b        &     -     &     -      &          & 144 (152) & 53 (79)  \\
\linea{Si}{iv}{4116} &  r        &  b        &     -     &     -      &          &   b       & 61 (69)  \\
\noalign{\smallskip}
\linea{O}{iii}{5592} &   -       & 213 (238) &     -     &      -     &    -     &    -      &  58 (80) \\
\noalign{\smallskip}
\linea{C}{iv}{5801}  &    -      &  b        &     -     &     -      &    -     &    -      &  50 (77) \\
\linea{C}{iv}{5811}  &    -      & 224 (235) &     -     &     -      &    -     &    -      &  50 (77) \\
\noalign{\smallskip}
\hline
\noalign{\smallskip}
\noalign{\smallskip}
\hline
\noalign{\smallskip}
\vsini\ (FT-He)  & \solu{81}{5}  &  \solu{204}{9}  & \solu{351}{7}  & \solu{308}{9}  & \solu{59}{8}  & \solu{146}{3}  & \solu{61}{13} \\
\vsini\ (FW-He)  & \solu{112}{9} &  \solu{221}{6}  & \solu{376}{26} & \solu{309}{10} & \solu{88}{15} & \solu{167}{10} & \solu{97}{17} \\
\vsini\ (FT-M)   &      84       &  \solu{219}{8}  &       -        & -              &     -         &  144           & \solu{54}{5} \\
\vsini\ (FW-M)   &      90       &  \solu{237}{2}  &       -        & -              &     -         &  152           & \solu{76}{4} \\
\noalign{\smallskip}
\hline
\noalign{\smallskip}
\end{tabular}
}
\normalsize 
\rm
\caption{\footnotesize Same as Table \ref{table3} but for mid-O 
dwarfs
\label{table6}
}
\end{center}
\end{table*}
%
\begin{table*}[!t]
\begin{center}
\tiny{
\begin{tabular}{c c c c c c c c}    
\hline
\hline
\noalign{\smallskip}
Star     & CygOB2-7 & CygOB2-8C & CygOB2-9 & CygOB2-11 & CygOB2-8A & HD\,210839 & HD\,192639 \\
SpT      & O3If     & O5If      & O5If     & O5If+     & O5.5I(f)  & O6I(n)fp   & O7Ib(f)    \\
Obs. run & WHT03    & WHT03     & WHT03    & WHT03     & WHT03     & INT92      & INT89      \\
\snr     & 250      & 250       & 150      & 200       & 350       & 150        & 220        \\
\noalign{\smallskip}
\hline
\noalign{\smallskip}
\linea{He}{i}{4387}   &    -          &    -          &    -          &     -         &     -        &    -      & 95 (121)  \\
\linea{He}{i}{4471}   &    -          & 180 (211)     & 97 (136)      & 107 (154)     & 87 (136)     & 223 (256) & 104 (163) \\
\linea{He}{i}{4713}   &    -          &    -          &    -          &     -         & 84 (102)     &    -      & 114 (125) \\
\linea{He}{i}{4922}   &    -          &    -          &    -          &     -         &     -        &    -      & 115 (130) \\
\linea{He}{i}{5875}   & 94 (117)      & 170 (205)     & 97 (119)      & 120 (134)     & 82 (125)     &    -      &     -     \\
\noalign{\smallskip}
\linea{C}{iii}{5695}  &    -          &     (249)$^a$ & 104 (120)$^a$ & 120 (140)$^a$ & 98 (118)$^a$ &    -      &     -     \\
\linea{C}{iv}{5801}   & 96 (100)$^a$  &    b          &    b          &      b        &     b        &    -      &     -     \\
\linea{C}{iv}{5811}   & 105 (103)$^a$ & 180 (200)     &    b          & 117 (143)     & 84 (120)     &    -      &     -     \\
\noalign{\smallskip}
\linea{O}{iii}{5592}  &     -         & 184 (196)     & 113 (135)     & 100 (136)     & 80 (135)     &    -      &     -     \\ 
\noalign{\smallskip}
\linea{Si}{iv}{4089}  & 89 (112)$^a$  &     -         &    -          &       -       &    -         &    -      & 100 (114) \\
\linea{Si}{iv}{4116}  & 89 (112)$^a$  &     (233)$^a$ &     (179)$^a$ & 122 (147)     &    (108)$^a$ &    -      & 117 (112) \\
\noalign{\smallskip}
\linea{N}{iii}{4634}  &  (120)$^a$    &     b         &     b         &     b         &     b        &     b     &     b     \\
\linea{N}{iii}{4641}  & 94 (120)$^a$  &     b         &     b         &     b         &     b        &     b     &     b     \\
\linea{N}{iv}{4058}   & 102 (143)     &     -         &     -         &     -         &     -        &     -     &     -     \\
\noalign{\smallskip}
\hline
\noalign{\smallskip}
\noalign{\smallskip}
\hline
\noalign{\smallskip}
\vsini\ (FT-He) & 94             & \solu{175}{7}  &       97       & \solu{114}{9}  & \solu{84}{3}   & 223 & \solu{107}{9}  \\
\vsini\ (FW-He) & 117            & \solu{208}{4}  & \solu{128}{12} & \solu{144}{14} & \solu{121}{17} & 256 & \solu{135}{19} \\
\vsini\ (FT-M)  & \solu{96}{7}   & \solu{182}{3}  & \solu{109}{6}  & \solu{115}{10} & \solu{87}{9}   &     & \solu{109}{12} \\
\vsini\ (FW-M)  & \solu{116}{14} & \solu{220}{26} & \solu{145}{31} & \solu{142}{5}  & \solu{120}{11} &     & \solu{113}{1}  \\
\noalign{\smallskip}
\hline
\noalign{\smallskip}
\end{tabular}
}
\normalsize 
\rm
\caption{\footnotesize Same as Table \ref{table3} but for early and mid-O 
supergiants. $^a$ Line in emission.
\label{table7}
}
\end{center}
\end{table*}
%
\subsection{Early and mid-O dwarfs and supergiants}
\label{section35}
%
When moving to earlier spectral types (mid and early O-type) most of the 
strong metal lines present in the optical spectra of early B-type stars
become very faint or even do not appear anymore. A few lines associated
with high ionized states of C, N, O, and Si can be found in mid O-type
stars, and a few metal lines in emission are present in the spectra of
the earliest spectral types. Contrarily, the spectra of mid O-type stars is
dominated by strong \ion{He}{i} and \ion{He}{ii} lines. And finally, 
even \ion{He}{i} lines become weak for the hottest stars. Therefore, in 
the optical spectra of those stars, only a few \ion{He}{} lines are 
available for the \vsini\ determination. \\

In previous sections, it has been shown that \ion{He}{i} lines result in
somewhat larger \vsini\ values than metal lines when measured through the 
\fwhm\ method. Contrarily, the \ft\ method produces 
an excellent agreement. Therefore, application of the \ft\ method to the
\ion{He}{i} lines in O-type stars is very promising to derive reliable
\vsini\ values.

It is important to note that \snr\ imposes an important limitation 
when the \ft\ method is applied to \ion{He}{i} lines. Since the Stark
broadening mechanism has an important influence on the \ion{He}{i} lines,
high \snr\ spectra are necessary to be sure that white noise in the 
Fourier space is not affecting the identification of the first rotational 
zero (as shown in Sect. \ref{section21}). The effect of noise may be even 
more dramatic if macroturbulence is also affecting the \ion{He}{i} line 
profiles. However, note that macroturbulent broadening is expected to be 
more important in giants and supergiants, while in these stars Stark 
effect contributes less to the broadening of the \ion{He}{i} lines (i.e. Stark 
broadening scales roughly as $\rho^2$). Note also that, although the 
spectra of early O-type stars show quite large \ion{He}{ii} lines, the 
effect of Stark broadening is larger for these lines than for \ion{He}{i} 
lines. Therefore, considering the comment above the use of these lines 
will require spectra with very high \snr. We did not achieve enough \snr\ 
in any of the spectra of our sample of early O-type stars, and hence 
\ion{He}{ii} could not be used.

Tables \ref{table6} and \ref{table7} summarize the line-to-line 
\vsini\ values resulting from the \fwhm\ and \ft\ analyses of the sample
of early and mid O-type dwarfs and supergiants. \ion{He}{i} lines were mainly
used; however, for the supergiants, we could also use some metal lines (some
of them present in the spectral range between 5000 and 6000 \AA, only observed
in the {\sc wht03} campaign). A comparison of \vsini\ results is again shown in
Figures \ref{figure5} to \ref{figure7} (dwarfs and supergiants
corresponding to black diamonds and red triangles, respectively).\\

From the comparison of \vsini\ values obtained for metal lines with the 
\ft\ and \fwhm\ methods in some of the supergiants in our sample, we can 
conclude 
that macroturbulence is also present in these stars, extending the result 
found in B-type supergiants to the O-type supergiants. Unfortunately, the 
presence of macroturbulence in the early O-type dwarfs could not be investigated 
because there was no metal lines available for the \vsini\ determination.
%
\subsection{OB giants}
\label{section36}
Results for the determination of projected rotational velocities in the 
sample of giant stars are presented in Tables \ref{table8} and \ref{table9}. 
Similar to previous sections, \vsini\ measurements from both the \ft\ and 
\fwhm\ for several lines, as well as mean values for \ion{He}{i} and metal 
lines, are summarized in those Tables. For these stars \ion{He}{i} lines 
were mainly used, although in some cases a few \ion{Si}{iv}, \ion{C}{iv}, 
\ion{N}{ii} and \ion{O}{iii} lines were also available.\\
\newline
The analysis of the sample of Giant stars points towards two curious results:
\begin{itemize}
\item{Both low and high \vsini\ values are found.}
\item{Macroturbulent effect is either very large or negligible.} 
\end{itemize}
These will be discussed in more detail in the next section.
%
\begin{table}[t]
\begin{center}
\tiny{
\begin{tabular}{c c c c c}    
\hline
\hline
\noalign{\smallskip}
Star     & CygOB2-8B & CygOB2-4 & HD164492A & HD\,168504  \\
SpT      & O6.5III   & O7III(f) & O7.5III   & O8III       \\
Obs. run & WHT03     & WHT03    & INT02     & WHT03       \\
\snr     & 230       & 220      & 250       & 300         \\
\noalign{\smallskip}
\hline
\noalign{\smallskip}
\linea{He}{i}{4387}   &     d    &  d      & 45 (89) & 88 (112)   \\
\linea{He}{i}{4471}   & 74 (130) & 83 (120 )   & 49 (94) & 79 (129)  \\
\linea{He}{i}{4713}   &          & 82 (118) & 52 (61) & 77 (90)        \\
\linea{He}{i}{4922}   & -        &  (128 )  & 52 (80) &           \\
\linea{He}{i}{5015}   & 94 (115) & 85 (116) & 48 (60) &            \\
\linea{He}{i}{5875}   & 85 (115) & 80 (130) &         &            \\
\noalign{\smallskip}
\linea{C}{iv}{5811}   & 83 (110) & 70 (136) &    -   &             \\
\noalign{\smallskip}
\linea{N}{ii}{4254}   & 65 (118) & 74 (125) &        &            \\
\noalign{\smallskip}
\linea{O}{iii}{5592}  & 82 (118) & 74 (125) &    -   &             \\
\noalign{\smallskip}
\linea{Si}{iv}{4089}  & 84 (119) & 84 (125) & 50 (61) &            \\
\linea{Si}{iv}{4116}  &          & 81 (103) & 45 (56) & 85 (89)    \\
\linea{Si}{iv}{4212}  &          &          & 50 (51) & 81 (81)    \\
\linea{Si}{iv}{4654}  &          &          & 45 (51) & 76 (81)    \\
\noalign{\smallskip}
\hline
\noalign{\smallskip}
\noalign{\smallskip}
\hline
\noalign{\smallskip}
\vsini\ (FT-He) & \solu{84}{10}  & \solu{83}{2}   & \solu{49}{3}  & \solu{81}{6} \\
\vsini\ (FW-He) & \solu{120}{9} & \solu{122}{6}  & \solu{77}{16} & \solu{110}{20} \\
\vsini\ (FT-M)  & \solu{79}{9}  & \solu{77}{6}   & \solu{48}{3}  & \solu{81}{5} \\
\vsini\ (FW-M)  & \solu{116}{4} & \solu{123}{12} & \solu{55}{5}  & \solu{84}{5} \\
\noalign{\smallskip}
\hline
\noalign{\smallskip}
\end{tabular}
}
\normalsize 
\rm
\caption{\footnotesize Same as Table \ref{table3} but for OB giants with 
         \vsini\ values below 100 \kms.
\label{table8}
}
\end{center}
\end{table}
%
\begin{table}[t]
\begin{center}
\tiny{
\begin{tabular}{c c c c}    
\hline
\hline
\noalign{\smallskip}
Star     & HD\,24912     & HD\,203064     & HD\,191423 \\
SpT      & O7III(n)((f)) & O7.5III:n((f)) & O9III:n \\
Obs. run & INT89         & INT89          & INT89 \\
\snr     & 250           & 200            & \\
\noalign{\smallskip}
\hline
\noalign{\smallskip}
\linea{He}{i}{4387}   & 224 (247) &           &           \\
\linea{He}{i}{4471}   & 211 (247) & 317 (338) &  468 (470) \\
\linea{He}{i}{4713}   & 225 (242) & 330 (322) &           \\
\linea{He}{i}{4922}   & 215 (230) & 328 (335) &  429 (437) \\
\noalign{\smallskip}
\hline
\noalign{\smallskip}
\noalign{\smallskip}
\hline
\noalign{\smallskip}
\vsini\ (FT-He) & \solu{219}{7} & \solu{325}{7} & \solu{450}{30} \\
\vsini\ (FW-He) & \solu{242}{8} & \solu{332}{9} & \solu{454}{23} \\
\noalign{\smallskip}
\hline
\noalign{\smallskip}
\end{tabular}
}
\normalsize 
\rm
\caption{\footnotesize Same as Table \ref{table3} but for OB giants with 
large \vsini.
\label{table9}
}
\end{center}
\end{table}
%
\section{Discussion}\label{section4}
%
Fig. \ref{figure5} shows a comparison of \vsini\ results obtained for the 
sample of stars when applying the \ft\ and \fwhm\ methods to metal and 
\ion{He}{i} lines. This comparison was only possible for stars with \vsini\ 
values below 200 \kms. For those cases with a projected rotational velocity 
above this value metal lines appeared blended or very faint, and hence they 
were not available to derive \vsini. For the studied stars, the \ft\ method 
produces coherent results between metal and \ion{He}{i} lines, indicating 
the possibility of accurately deriving projected rotational velocities through 
this method by using \ion{He}{i} lines. This is not the case for the \fwhm\ 
method, whose \ion{He}{i} lines \vsini\ associated values are somewhat 
larger than those derived from metal lines, even for large values of \vsini\ 
(see also Fig. \ref{figure6}).
This result is a consequence of the fact that \ion{He}{i}
lines are affected by Stark broadening, while we are considering a Gaussian
definition of the intrinsic profiles. It indicates that the assumption made 
by Herrero et al. (\cite{Her92}), that this hypothesis can be used
when applying the \fwhm\ method to \ion{He}{i} lines for determining \vsini\ 
if this is large, is not completely correct. The effect of pressure broadening 
over \ion{He}{i} may be important even for large projected rotational 
velocities, although for \vsini\,$>$\,250 km s$^{-1}$ the relative error 
is small. Note that, although this discrepancy should not be present 
if the intrinsic profiles used for the \ion{He}{i} lines consider the 
Stark broadening, the derived \fwhm\ values may require a large computational 
effort and be model dependent , while \ft\ values will not.

Since the Stark effect over the \ion{He}{i} lines scales roughly as $\rho^2$,
these lines are expected to be less affected by Stark broadening 
in supergiants. This explains why for some of the studied supergiants,
a better \ion{He}{i} vs. metal lines agreement is found when applying
the \fwhm\ method (Fig. \ref{figure5}, right). However, caution!, the
agreement between \ion{He}{i} and metal lines does not necessarily 
mean that the \fwhm\ method is giving correct values for the \vsini.\\

Differences between the \vsini\ values obtained through the application of
the \fwhm\ and \ft\ methods to metal lines can be interpreted as an indicator 
of the presence of some extra broadening, traditionally associated to large 
scale motions. As indicated above, we will refer to it as macroturbulence.
This is illustrated in Fig. \ref{figure7}. The whole sample of stars whose 
metal lines are available for the \vsini\ to be measured are considered 
in this figure. Note that a similar comparison with \ion{He}{i} lines (i.e.
Fig. \ref{figure6}) does 
not have sense for investigating the presence of macroturbulence using 
the differences between the \ft\ and \fwhm\ methods since these lines 
are also affected by other non-rotational broadenings as shown above
(metal lines are also affected by other line broadening mechanisms,
like thermal or collisional ones, but we expect them to be negligible, as
shown by the agreement found in dwarf analyses). Stars have been separated 
according to luminosity class.
While macroturbulence can be considered negligible in dwarfs, supergiants are 
clearly affected by this extra-broadening mechanism; 65\% of the sample of 
supergiants shows \vsini\,(\ft)\,$\le$\ 0.8 \vsini\,(\fwhm). The case of 
giants is more complicated; these objects display a large range of 
\vsini (\ft)/\vsini (\fwhm) ratios. 
%
\begin{figure}[t!]
\centering
\includegraphics[width=6.8cm,angle=90]
{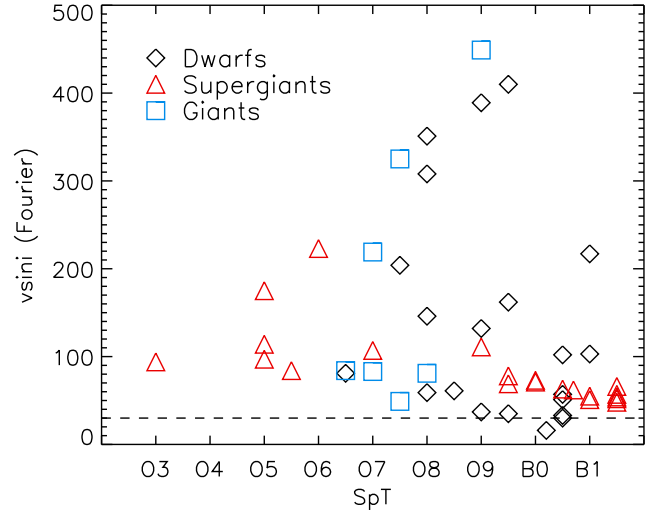}
\caption{Distribution of projected rotational velocities (as determined by
the \ft\ method) among the various spectral types of the stars in the sample.
Both \vsini$_{\rm FT}$(metal) and \vsini$_{\rm FT}$(\ion{He}{i}) are used 
in this figure. Horizontal dashed line represent the resolution limit for
R\,=\,9000.}
\label{figure9}
\end{figure}

Supergiants rotate as a group more slowly than dwarfs do. Except for two 
cases among early O-type supergiants (CygOB2-8C and HD\,218039) all of 
them have projected rotational velocities below or very close to 100 \kms. 
This has to be compared with the large range of \vsini\ found for dwarfs 
and giants, that cover from about 15 up to 450 \kms. In Fig. \ref{figure9} 
we see that the region of low rotational velocities is shared by all spectral 
types and luminosity classes, but that of large rotational velocities is 
dominated by giants and dwarfs of mid and late spectral types. This 
result was already indicated by Slettebak (\cite{Sle56}) and later on by 
other authors (viz. Conti \& Ebbets \cite{Con77}, Howarth et al. \cite{How97}).
These authors also propose indirect indications of the possible presence 
of macroturbulence in OB stars, but could not directly confirm the amount 
of broadening that is actually due to rotation and therefore
their \vsini values are affected by the contribution of macroturbulence.
The \ft\ method shows that indeed very large rotational velocities 
are the cause for the large line broadening in dwarfs, and allows 
for a quantitative estimation of the actual \vsini\ in supergiants 
(where effects of macroturbulence are important). First results for
a more limited sample were presented by Ebbets (\cite{Ebb79}). Our 
results also 
support the bimodality found by Conti and Ebbets (\cite{Con77}) and 
other authors since then in the \vsini\ distribution of main sequence 
stars.\\

Although it is clear that supergiants are affected by macroturbulence,
our knowledge of its behavior and the physical mechanisms which 
produce this extra broadening is still very poor. In Sect. \ref{section21}
an isotropic Gaussian macroturbulence was assumed for the formal tests;
however, as shown by Dufton et al. (\cite{Duf06}), one cannot rule out
other macroturbulent velocity distributions from the comparison with 
high resolution observed line profiles. Physically, a 
Gaussian-angle-dependent velocity distribution similar to that proposed
by Gray (\cite{Gra76}) to model the flow of convective cells in the Sun,
may also have sense in the case of B supergiants. In that case, the 
large-scale radial-tangential movements may be produced when the stellar 
material flowing outwards as a stellar wind is coupled with the stellar 
rotation.
Observational tests to investigate the exact degree of the anisotropy of
the macroturbulence should be investigated in the future. In the meantime, rough 
estimations of the macroturbulence are also valuable, once the projected
rotational velocity has been inferred. Note that the characteristic value
of the macroturbulent velocity is dependent on the approach considered
for the macroturbulent velocity distribution (see p.e. Dufton et al. 
\cite{Duf06}).\\

\begin{figure}[t!]
\centering
\includegraphics[width=7.0cm,angle=90]
{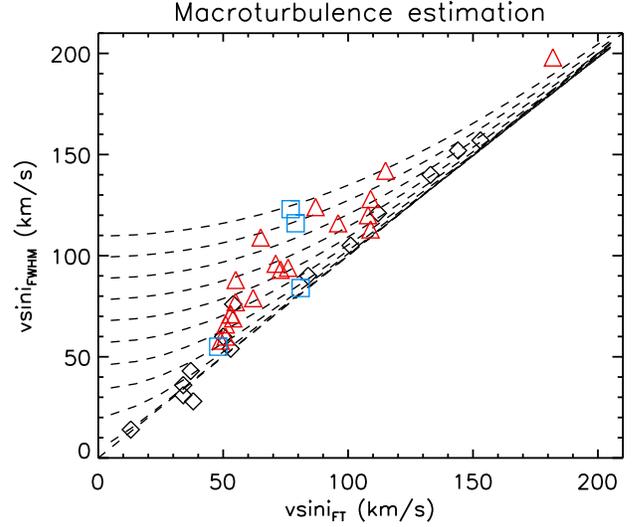}
\caption{This plot illustrates the method used to derive $\Theta_0$, a 
rough estimation of the macroturbulence. The various dashed curves 
represent cases with different values of $\Theta_0$, from 5 to 95
\kms. An instrumental profile (R\,=\,9000) has also been considered
(see text for explanation). Symbols have the same meaning than in 
previous figures.}
\label{figure10}
\end{figure}

The quality of our observational data-set (mainly selected to show
the applicability of the \ft\ method to OB stars) is not good enough to
proceed to a fine estimation of the macroturbulence\footnote{We refer to 
the works by Dufton et al. (\cite{Duf06}) and Lefever et al. (\cite{Lef06})
as examples of macroturbulent determination through better quality 
observational data.}. However, taking into account the discussion above,
we have decided to consider a rough estimation of the amount of 
extra-broadening, given by $\Theta_0$, where this quantity is calculated 
as follows. A \vsini\,-\fwhm\ calibration is created by measuring the \fwhm\ 
corresponding to the convolution of an instrumental profile and various 
values of \vsini. This calibration is applied to a second set of line 
profiles in which an isotropic Gaussian macroturbulence (with characteristic 
velocity $\Theta_0$) is also included.
The corresponding \vsini\,$_{\rm FWHM}$ is hence inferred for various 
\vsini\,-\,$\Theta_0$ combinations, ending with a set of lines in the 
\vsini\,$_{\rm FWHM}$\,-\,\vsini\,$_{\rm FT}$ plane (see Fig. 
\ref{figure10}). Once this two quantities are measured for a given
star, the $\Theta_0$ can be estimated by interpolation between these curves.
We expect the quantity $\Theta_0$ calculated in this way to be roughly 
proportional to the macroturbulent velocity and behave in a similar way 
when varying the stellar parameters.

\begin{figure}[t!]
\centering
\includegraphics[width=6.8cm,angle=90]
{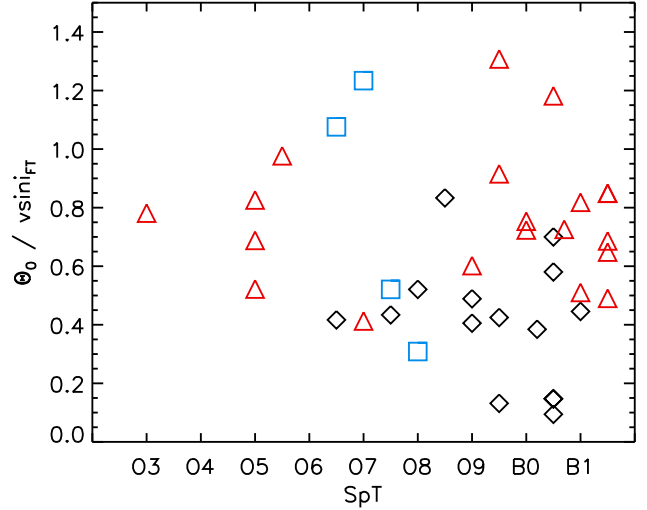}
\caption{Distribution of the ratio $\Theta_{\rm 0}$/\vsini\ in 
the sample of stars as a function of spectral types. Only those stars 
with metal lines available to measure \vsini$_{\rm FW}$\ and 
\vsini$_{\rm FT}$ were used (see text and Figure \ref{figure10}).}
\label{figure11}
\end{figure}

The influence of $\Theta_0$ relative to \vsini\ seems to increase 
when moving from dwarfs to supergiants, as can be seen in Fig. \ref{figure11}.
We have decided to plot $\Theta_0$/\vsini\ vs. SpT, instead of $\Theta_0$,
because for large values of \vsini\ is more difficult to distinguish between
the different values of $\Theta_0$, as it is illustrated by the proximity
of the dashed curves when \vsini\ increases. We consider this ratio more
meaningful. 
            
Except for the two giants around O7, the largest values at each spectral 
type are always obtained for supergiants. Only two dwarfs have 
$\Theta_0$/\vsini\ comparable 
to the one found in Supergiants; these are HD\,37023, which is known to
be binary, and HD\,168137 (O8.5V).
%
%

\begin{figure}[t!]
\centering
\includegraphics[width=6.2cm,angle=90]
{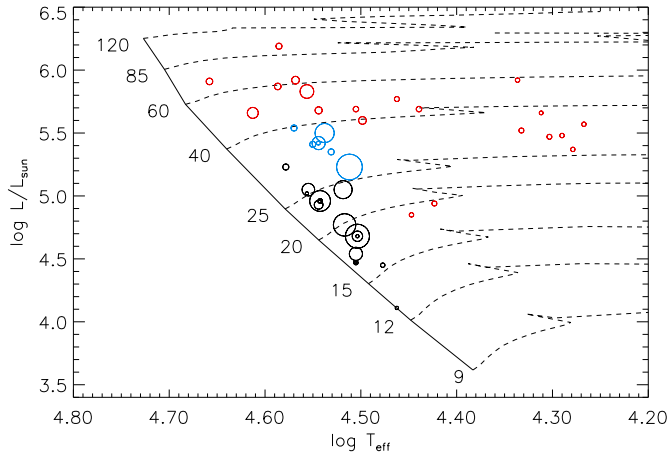}
\caption{HR diagram where the stars in the studied sample have been
located with symbol sizes proportional to their rotational (\vsini$_{\rm FT}$) 
velocities. As in previous figures, black, blue, and red symbols represent
dwarfs, giants and supergiants, respectively.}
\label{figure12}
\end{figure}

\begin{figure}[t!]
\centering
\includegraphics[width=6.2cm,angle=90]
{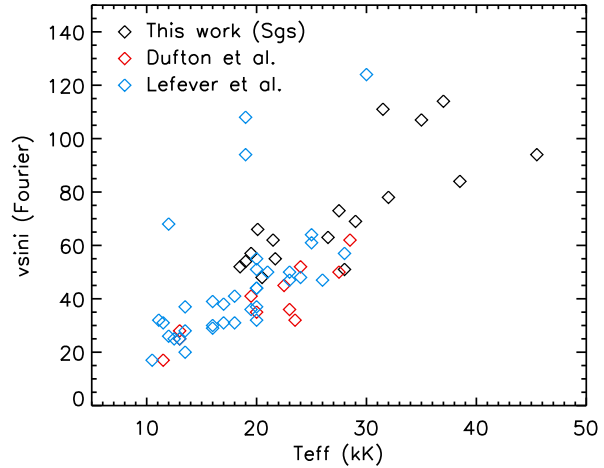}
\caption{Estimates of projected rotational velocities for our sample of
Supergiant stars. Results from the analysis of the Galactic B-type 
supergiants by Dufton et al. and Lefever et al. are also included.}
\label{figure13}
\end{figure}

Combining both results, we conclude that the role of rotation should 
decrease and the role of macroturbulence increase when the stars move from 
dwarfs to supergiants, i.e, when they evolve. This can also be seen in
Figures \ref{figure12} and \ref{figure13}, where we have represented the
stars on a HR diagram,with symbol sizes proportional to their rotational
and macroturbulent velocities. To place the stars in the HRD we have
taken results from various authors that used state-of-the-art {\sc nlte}, 
line-blanketed, unified model atmospheres (Herrero et al. \cite{Her02}, 
Repolust et al. \cite{Rep04}, Urbaneja \cite{Urb04}, Sim\'on-D\'iaz 
\cite{Sim05}, Sim\'on-D\'iaz et al. \cite{Sim06}, Crowther et al. 
\cite{Cro06}). For those stars in the sample which have not been previously 
analyzed, we use the SpT-\Teff\ and SpT-$L$ calibrations defined by Martins 
et al. (\cite{Mar05}). Note that for some of the B0.5V stars, the stellar 
parameters were obtained by extrapolating the calibration by Martins et al.

We see that the symbols in Fig. \ref{figure12} may have any size close 
to the ZAMS, but they become smaller when we depart from it. Beyond an 
imaginary line parallel to the ZAMS all symbols are very small, indicating 
slow rotation. Although the inclination angle is unknown, it is obvious
that the effect cannot be attributed to a random effect. Clearly, all
supergiants have small rotational velocities which, in  view of the
distribution for dwarfs and giants, indicates that the stellar rotational
velocities decrease when the stars evolve. As suggested by other authors 
(see review by Herrero \& Lennon \cite{Her04}) this may be related to the 
loss of angular momentum caused by stellar winds.\\
%

\begin{figure}[t!]
\centering
\includegraphics[width=6.cm,angle=90]
{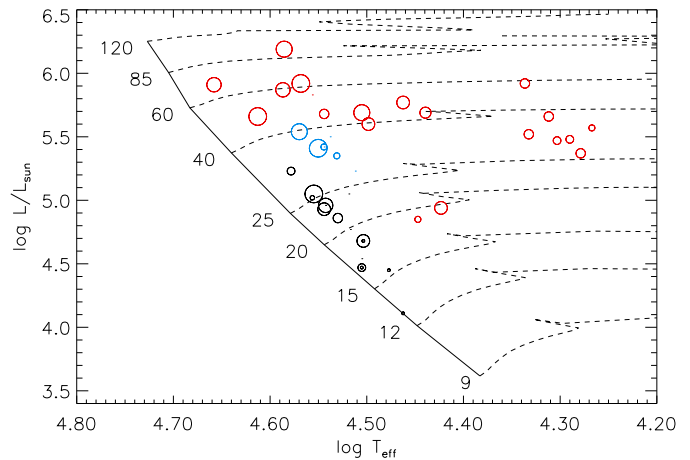}
\caption{Same as in Fig. \ref{figure12} but for the quantity 
$\Theta_{\rm 0}$ (i.e. rough estimation of the macroturbulence)}
\label{figure14}
\end{figure}

Recently Dufton et al. (\cite{Duf06}) and Lefever et al. (2006)
have applied the \ft\ method to two samples of Galactic B-type supergiants 
observed with high spectral resolution. They found larger \vsini\ values 
for the earliest spectral types. We put together the \vsini\ results by 
these authors and our results in Fig. \ref{figure14}. We can see that 
they agree very well, and that our results extend those of Dufton et al. 
and Lefever et al. towards higher temperatures.

Additionally, these authors could use the good quality of their data to 
estimate the amount of macroturbulent broadening present in the stars 
by means of the fit of the theoretical line profiles to the observed 
ones. Since we limit ourselves to a crude first order approximation as 
shown above, we have to be very careful in interpreting the results 
concerning our estimation of what we have called 
macroturbulence\footnote{Note that even a comparison of the values 
inferred by these authors for the macroturbulence should be handle with 
care, since Dufton et al. are using an isotropic Gaussian definition while 
Lefever et al. assume a radial-tangential approach.}.
However, it is interesting to see how this estimation varies across the 
HR diagram. In Fig. \ref{figure13} we see that the symbol sizes representing 
the macroturbulent velocity have a much more uniform distribution than 
those of the rotational velocities, with the lowest values corresponding 
to dwarfs close to the ZAMS.
Part of this difference may be related to projection effects (that would 
affect large scale turbulent motions much less than rotation). 
Nevertheless, according to these estimations, macroturbulence
does not necessarily increase significantly when the stars evolve
redwards, but may remain approximately constant. However, its role 
in line broadening relative to rotation increases (and perhaps also in 
other physical mechanisms). Note that this broad scenario suggested here 
(rotational velocity decreasing when the stars evolve from dwarfs to 
supergiants, macroturbulence being roughly constant but taking over a 
dominant role) has been deduced from the study of a very inhomogeneous 
sample of OB stars. To better understand the role that rotation and
macroturbulence play in the evolution of massive stars, an homogeneous
study of stars in clusters of different ages and metallicities should be
performed (viz. Dufton et al. \cite{Duf06b}, Mokiem et al. \cite{Mok06}).
%
%
%
\begin{table*}[!h]
\begin{center}
\scriptsize{
\begin{tabular}{c c c c c c c}    
\hline
\hline
\noalign{\smallskip}
Star	& SpT	& \vsini (\ft) & \vsini(\fwhm) & \Teff (K) & log\,$L$/$L_{\odot}$ & Ref \\
\noalign{\smallskip}
\hline
\noalign{\smallskip}
CygOB2-7 	& O3If* 	& \solu{96}{7}   & \solu{116}{14} & 45500 & 5.91 & He02\\
CygOB2-8C 	& O5I(f) 	& \solu{182}{3}  & \solu{220}{26} & 41000 & 5.66 & He02\\
CygOB2-9 	& O5If+ 	& \solu{109}{6}  & \solu{145}{31} & 38610 & 5.87 & Mar05 \\
CygOB2-11 	& O5I(f)+ 	& \solu{115}{10} & \solu{142}{5}  & 37000 & 5.92 & He02 \\
CygOB2-8A 	& O5.5If 	& \solu{87}{9}   & \solu{124}{9}  & 38500 & 6.19 & He02 \\
HD\,210839 	& O6I(n)fp 	&  223$^{(a)}$   &    256$^{(a)}$ & 36000 & 5.83 & Rep04 \\
HD\,192639 	& O7Ib(f) 	& \solu{109}{12} & \solu{113}{1}  & 35000 & 5.68 & Rep04 \\
HD\,210809 	& O9Iab 	& \solu{108}{4}  & \solu{120}{7}  & 31500 & 5.60 & Rep04\\
CygOB2-10 	& O9.5Iab 	& \solu{71}{9}   & \solu{95}{12}  & 29000 & 5.77 & He02\\
HD\,209975 	& O9.5Iab 	& \solu{66}{6}   & \solu{103}{10} & 32000 & 5.69 & Rep04\\
HD\,167264 	& B0Ia 	    & \solu{73}{4}   & \solu{93}{11}  &       &      & \\
HD\,37128 	& B0Ia 	    & \solu{76}{7}   & \solu{94}{8}   & 27500 & 5.69 & Urb \\
HD\,38771 	& B0.5Ia 	& \solu{55}{5}   & \solu{88}{7}   & 26500 & 4.94 & Urb\\
HD\,2905 	& BC0.7Ia 	& \solu{62}{6}   & \solu{79}{6}   & 21500 & 5.52 & Cro06 \\
CygOB2-2 	& B1I 	 	& \solu{49}{9}   & \solu{58}{8}   & 28000 & 4.85 & He02\\
HD\,13854 	& B1Iab 	& \solu{55}{6}   & \solu{77}{6}   & 21700 & 5.92 & Urb\\
HD\,190603 	& B1.5Ia+ 	& \solu{51}{4}   & \solu{60}{5}   & 18500 & 5.57 & Cro06 \\
HD\,14956 	& B1.5Ia 	& \solu{53}{5}   & \solu{71}{4}   & 20500 & 5.66 & Urb\\
HD\,14143 	& B2Ia 	 	& \solu{51}{6}   & \solu{66}{6}   & 19500 & 5.48 & Urb\\
HD\,14818 	& B2Ia 	 	& \solu{54}{6}   & \solu{69}{6}   & 20100 & 5.47 & Urb\\
HD\,194279 	& B2Ia 	 	& \solu{53}{3}   & \solu{71}{3}   & 19000 & 5.37 & Cro06 \\
\noalign{\smallskip} 
\hline
\noalign{\smallskip}
CygOB2-8B	    & O6.5III(f)     & \solu{79}{9}           & \solu{116}{4}          & 37130 & 5.54 &  Mar05 \\
HD\,24912	    & O7III(n)((f))	 & \solu{219}{7}$^{(a)}$  & \solu{242}{8}$^{(a)}$  & 35000 & 5.42 &  Rep04 \\
CygOB2-4	    & O7III(f)	     & \solu{77}{6}           & \solu{123}{12}         & 35500 & 5.41 &  He02 \\
HD\,164492A	    & O7.5III	     & \solu{48}{3}           & \solu{55}{5}           & 35020 & 5.42 &  Mar05 \\
HD\,203064	    & O7.5III:n((f)) & \solu{325}{7}$^{(a)}$  & \solu{332}{9}$^{(a)}$  & 34500 & 5.50 &  Rep04 \\
HD\,168504      & O8III          & \solu{81}{5}           & \solu{84}{5}           & 33960 & 5.35 &  Mar05 \\
HD\,191423      & O9III:n        & \solu{450}{30}$^{(a)}$ & \solu{454}{23}$^{(a)}$ & 32500 & 5.23 &  Rep04 \\
\noalign{\smallskip}
\hline
\noalign{\smallskip}
HD\,12993	    & O6.5V	   &      84                &        90              & 37870 & 5.23 &  Mar05 \\
CygOB2-16	    & O7.5V	   & \solu{219}{8}          & \solu{237}{2}          & 35870 & 5.05 &  Mar05 \\
HD\,46056	    & O8V((f)) & \solu{351}{7}$^{(a)}$  & \solu{380}{30}$^{(a)}$ & 34980 & 4.96 &  Mar05 \\
HD\,13268	    & ON8V	   & \solu{308}{9}$^{(a)}$  & \solu{309}{10}$^{(a)}$ & 33000 & 5.05 &        \\
HD\,46966	    & O8V	   & \solu{59}{8 }$^{(a)}$  & \solu{88}{14}$^{(a)}$  & 34880 & 4.96 &  Mar05 \\
HD\,236894      & O8V	   &        144             &          152           & 34880 & 4.96 &  Mar05 \\
HD\,168137      & O8.5V	   & \solu{54}{5}           & \solu{76}{4}           &       &      &  Mar05 \\
HD\,37041	    & O9V	   & \solu{132}{5}$^{(a)}$  & \solu{165}{21}         & 35000 & 4.93 &  SD06  \\
HD\,214680	    & O9V	   & \solu{37}{3}           & \solu{43}{4}           & 36000 & 5.02 &  SD06  \\
HD\,149757	    & O9V	   & \solu{390}{17}$^{(a)}$ & \solu{414}{18}$^{(a)}$ & 32880 & 4.77 &  Mar05 \\
BD-134928       & O9.5Vn   & \solu{410}{13}$^{(a)}$ & \solu{420}{8}$^{(a)}$  & 31880 & 4.68 &  Mar05 \\
CygOB2-23	    & O9.5V	   & \solu{162}{8}$^{(a)}$  & \solu{177}{20}$^{(a)}$ & 31880 & 4.68 &  Mar05 \\
CygOB2-145	    & O9.5V	   & \solu{38}{4}           & \solu{28}{8}           & 31880 & 4.68 &  Mar05 \\
$\tau$\,Sco     & B0.2V	   & $\le$\,13              & \solu{14}{3}           & 32000 & 4.47 &  SD06  \\
HD\,37020	    & B0.5V	   & \solu{53}{3}           & \solu{54}{3}           & 30000 & 4.45 &  SD06  \\
HD\,37023	    & B0.5V	   & \solu{50}{3}           & \solu{60}{8}           & 32000 & 4.47 &  SD06  \\
HD\,37042	    & B0.5V	   & \solu{34}{3}           & \solu{36}{4}           & 29000 & 4.11 &  SD06  \\
HD\,217657	    & B0.5V	   & \solu{34}{5}           & \solu{31}{6}           &       &      & \\
HD\,228199	    & B0.5V	   & \solu{112}{8}          & \solu{121}{6}          &       &      & \\
HD\,37061	    & B1V	   & \solu{217}{3}$^{(a)}$  & \solu{231}{23}$^{(a)}$ & 32000 & 4.54 &  SD05 \\
$\omega^1$\,Sco & B1V      & \solu{101}{4}          & \solu{105}{5}          &       &      & \\
\noalign{\smallskip}
\hline
\hline
\\
\end{tabular}
}
\normalsize 
\rm
\caption{\footnotesize All \vsini\ values correspond to metal lines 
measurements except those marked with $^{(a)}$. References for the effective 
temperatures and luminosities: He02 (Herrero et al. 2002), Rep04 (Repolust et 
al. 2004), Urb04 (Urbaneja, 2004, thesis) , Cro06 (Crowther et al. 2006), SD05 
(Sim\'on-D\'iaz, 2005, thesis) SD06 (Sim\'on-D\'iaz et al. 2006). For the 
stars in the sample which have not been previously analyzed with line-blanketed 
models, we use the 
SpT\,-\,\Teff\ and SpT\,-\,$L$ calibrations defined by Martins et al. 
(2005, Mar05).
\label{table10}
}
\end{center}
\end{table*}

\section{Conclusions}
\label{section5}
%
In the past the Fourier method for the determination of projected rotational 
velocities has been extensively used in the analysis of late-type stars, but 
only marginally applied to the early-type case. In this paper we show the 
strength of this method for its application to OB stars.\\

The noise and the spectral dispersion of the observed spectra are the main 
limitations to the applicability of the Fourier method. Spectral dispersions 
below 0.4 \AA/pix are required to measure \vsini\ values below 30\,-\,40 \kms\
when using the optical spectrum of OB stars. For dwarf stars, not very
affected by non-rotational broadenings (i.e. macroturbulence), the \ft\
method can be successfully applied even for \snr\ values as low as 100. 
The effect of noise is more important in the case of supergiants, where
absorption stellar lines are broadened by mechanisms other 
than rotation. If the \snr\ of the observed spectra is not high enough
and the amount of macroturbulent broadening is of the order of rotational
broadening, the application of the \ft\ may result in somewhat larger \vsini\
values than the actual ones; \snr\ values above 300\,-\,400 are required
for these cases.\\

The comparison of \vsini\ values obtained through the \ft\ and \fwhm\ 
methods in a sample of O and early-B stars shows that the \fwhm\ technique 
must be used with care in the analysis of OB giants and supergiants, and 
when it is applied to \ion{He}{i} lines. Contrarily, the \ft\ method 
appears to be a powerful tool to derive reliable projected rotational 
velocities, and separate the effect of rotation from other broadening 
mechanisms present in these stars.\\

The analysis of the sample of OB stars, that includes dwarfs, giants and 
supergiants with spectral types earlier than B2, shows that while dwarfs 
display a broad range of projected rotational velocities, from less than 
30 up to 450 km s$^{-1}$, supergiants have in general values close to or 
below 100 km s$^{-1}$. Giants also display a large range of projected 
rotational velocities. The analysis has also definitely shown that while 
the effect of macroturbulence in OB dwarfs is usually negligible when compared 
with the rotational broadening, the effect of this extra-broadening is 
clearly present in supergiants. \\

When examining the behavior of the projected rotational velocities with
the stellar parameters and across the HR diagram, we conclude, in agreement
with previous researchers, that the rotational velocity should decrease when
the stars evolve. On the contrary, extra broadening across the HR diagram, 
interpreted as macroturbulence, may be constant, resulting therefore in an 
increasing importance as compared to rotation when the stars evolve. 
%
\begin{acknowledgements}
We thank C. Trundle for her detailed reading of the first manuscript 
of this work, and for many fruitful discussions and comments.
We also thank M.R. Villamariz, J. Puls, D. J. Lennon and P. Dufton 
for numerous insightful discussions. We acknowledge the comments by 
the anonymous referee. 
This work has been partially funded by the Spanish Ministerio 
de Educaci\'on y Ciencia under project AYA2004-08271-C02-01.
S.-D. also acknowledge the funds by the Spanish Ministerio de 
Educaci\'on y  Ciencia under the MEC/Fullbright postdoctoral 
fellowship program. This paper is based on observations made 
with the Isaac Newton and William Herschel telescopes, operated 
on the island of La Palma by the Isaac Newton Group in the Spanish
Observatorio del Roque de los Muchachos of the Instituto de 
Astrof\'isica de Canarias.
\end{acknowledgements}
%
%

%
\end{document}